# CellPrior-Net: Prior-Guided Nuclei Detection and Classification for H&E Whole-Slide Images

**Falah Jabar[1*], Pasquale Lombardi[2], Aria Torkpour[2], Masoud Tafavvoghi[3], Per Niklas Benzler Waaler[3], Sigve Andersen[4], Erna-Elise Paulsen[5], Mette Pøhl[6], Lill-Tove Rasmussen Busund[1], Tom Donnem[4], Elin Richardsen[1], David J. Pinato[2], Mehrdad Rakaee[7]**

[1] Dep. of Clinical Pathology, University Hospital of North Norway, Tromsø, Norway

[2] Dep. of Surgery and Cancer, Imperial College London, London, United Kingdom

[3] Department of Medical Biology, UiT The Arctic University of Norway, Tromsø, Norway

[4] Dep. of Clinical Medicine, UiT The Arctic University of Norway, Tromsø, Norway

[5] Dep. of Pulmonology, University Hospital of North Norway, Tromsø, Norway

[6] Dep. of Clinical Medicine, University of Copenhagen, Copenhagen

[7] Dep. of Cancer Genetics, Oslo University Hospital, Oslo, Norway

*Corresponding author:

Falah Jabar, PhD
Hansine Hansens veg 67, 9019 Tromsø, Norway
E-mail: f.rahim@imperial.ac.uk

**Abstract**—Accurate nuclei detection and classification in hematoxylin and eosin (H&E) whole-slide images (WSIs) is a key task in computational pathology, particularly for quantitative analysis of the tumor microenvironment. However, this task remains highly challenging due to variations in nuclei morphology, staining procedures, scanners, organs, magnifications, and WSI artifacts. In addition, many existing pipelines rely on computationally demanding architectures and post-processing procedures, making gigapixel WSI analysis time-consuming. In this work, CellPrior-Net (CP-Net) is proposed, an efficient nuclei detection and classification pipeline that utilizes a lightweight convolutional neural network architecture and hematoxylin (H) channel as prior information to enhance nuclei-aware feature learning. Extensive benchmarking was conducted against state-of-the-art pipelines on 8 public and private datasets (total:~10.4M nuclei) obtained from different organs, scanners, magnifications, and clinical centers. Experimental results demonstrate that CP-Net achieves comparable performance while significantly reducing inference time. Furthermore, CellQuant-Net was introduced–an end to end nuclei quantification pipeline–that integrates a quality assessment (QA) model to exclude regions with artifacts, followed by CP-Net cell detection and classification. The pipeline is publicly available on GitHub, and provides a potentially efficient and scalable framework for downstream computational pathology applications.

***Keywords*:** Computational pathology, Computer vision, Digital pathology, Deep learning, Nuclei detection and classification, H&E Whole-slide images

## I. Introduction

Whole-slide imaging enables the digitization of histopathology slides into gigapixel WSIs, providing detailed histological information. Since WSIs are extremely large and cannot be directly processed by deep learning (DL) models, they are divided into smaller image tiles (**Fig. 1**). These tiles are subsequently used as input for various DL tasks, such as quantification, histological subtyping, genomic alterations, and treatment outcome prediction [1][2][3].
Several pathology foundation models have been developed for computational pathology applications [4][5][6][7][8][9][10][11][12][13][14][15][16]. In particular, nuclei detection and classification have become a primary task, as they enable quantitative analysis of the tumor microenvironment and immune infiltration. Thus, several DL pipelines for nuclei detection and classification have been proposed [17][18][19][20][21][22][23][24][25].

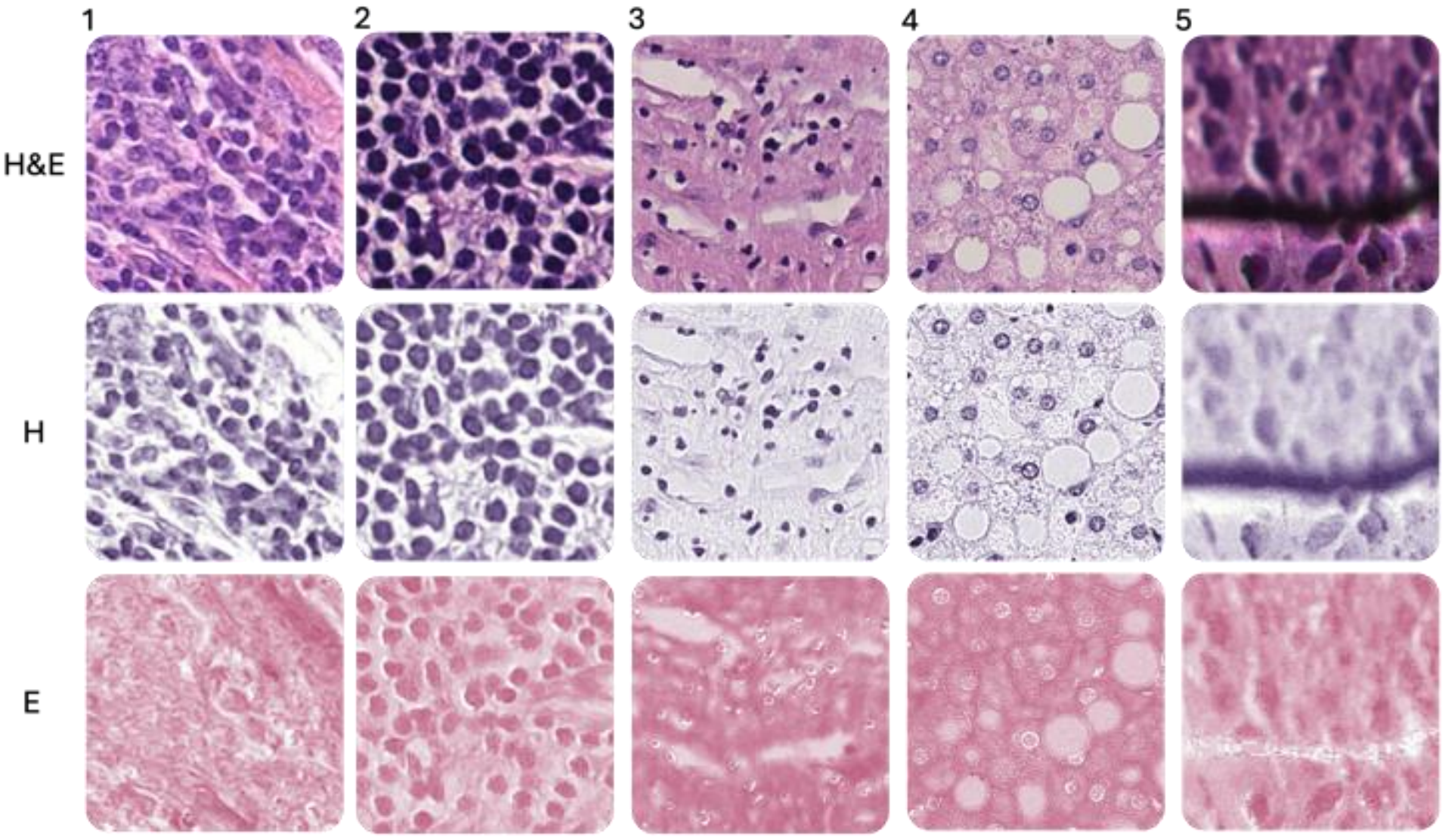

**Fig. 1**. Examples showing the variability in nuclei appearance in H&E WSIs. Columns 1–5 present different tissue and nuclei patterns extracted from multiple WSIs. The first row shows the original H&E image tile, while the second and third rows show the separated hematoxylin (H) and eosin (E).

However, up to date, accurate nuclei detection and classification remain highly challenging due to several factors (**Fig. 1**):

- **Morphology:** Differences in nuclei shape, size, texture, and appearance across tissues and cancer grades.
- **Organs:** Different organs have diverse tissue architectures, cell distributions, and appearances.
- **Staining:** Differences in staining procedures between laboratories introduce color and contrast inconsistencies.
- **Slicing:** During tissue slicing, some nuclei may appear partially visible or hidden beneath the tissue slice.
- **Scanner:** Different scanners produce variations in resolution, color distribution, sharpness, and illumination.
- **Overlapping**: Separating touching nuclei is challenging, particularly in regions with unclear nuclei boundaries.
- **Artifacts:** Artifacts (e.g., folds, blur, pen markings) may partially or completely obscure nuclei regions.
- **Magnification:** Nuclei may appear with different sizes and appearances at 20x versus 40x magnification.
- **Gigapixel WSI:** Processing gigapixel WSIs is computationally expensive, especially for nuclei detection and classification.
- **Dataset Annotation:** Nuclei annotation is highly challenging and time-consuming due to manual labeling.

Some of these challenges introduce domain shifts that cause the model to fail frequently on unseen data [19]. Also, most of the existing pipelines require considerable computational resources to process a single WSI due to complex model architecture and computationally expensive post-processing procedures. For example, processing a single high-resolution WSI may take several hours, or even days in some cases [25]. Moreover, it remains unclear how well existing nuclei detection and classification pipelines generalize across different datasets, organs, scanners, magnifications, and clinical centers. Many pipelines rely on post-processing to separate touching nuclei using predicted nuclei edge masks and may often fail for accurate classification at 20x magnification, as nuclei appear denser and more clustered (**Fig. 3**). Furthermore, none of these pipelines consider quality assessment (QA) to exclude regions with artifacts before or during nuclei detection and classification. As shown in our previous work [26] and related work [27][28], artifacts can significantly reduce nuclei detection and classification accuracy.

Together, these limitations highlight the need for WSI-scale nuclei analysis pipelines that integrate quality assessment, accurate nuclei detection and classification, and computational efficiency. This is particularly important because nuclei analysis requires dense instance-level processing of gigapixel WSIs, while most existing pipelines operate directly on full H&E images despite nuclei being predominantly represented in the hematoxylin component. Separating H&E components can be beneficial as the H channel enhances nuclei regions while reducing background noises (**Fig. 1**). In this context, the main objective of this paper is to develop a lightweight nuclei detection and classification pipeline, CP-Net, which utilizes the hematoxylin (H) channel as prior information to guide the network toward nuclei-relevant features, while using an efficient post-processing procedure that generalizes across different magnifications.

The main contributions of this work are:

- Development of the ILCD dataset, collected from 40 liver cancer WSIs at Imperial College London, annotated manually by experts pathologists. The dataset contains three cell categories: immune, tumor, and other cells (detailed in Supplementary Material).
- Development of a lightweight nuclei detection and classification pipeline that utilizes both H&E images and the H channel as prior information to classify nuclei into three categories: immune, tumor, and other cells (**Fig. 2**).
- Extensive benchmarking experiments against state-of-the-art pipelines using 8 public and private datasets collected from different organs, scanners, magnifications, and both single-organ and multi-organ WSIs (detailed in Supplementary Material).
- Integration of a QA pipeline with CP-Net to first remove artifacts from WSIs before nuclei detection and classification, followed by a nuclei connectivity analysis to model nuclei spatial relationships (**Fig. *5***).

The rest of this paper is organized as follows: Section II reviews related work, Section III describes the datasets, Section IV presents the proposed CP-Net methodology, Section V reports the performance evaluation results, Section VI introduces CellQuant-Net, and Section VII concludes the paper and discusses future directions.

## II. Related Work

Nuclei detection and classification pipelines have evolved from convolutional neural network (CNN)-based to vision transformer (ViT)-based and foundation model (FM)-based approaches. In [22], HoVer-Net was introduced as one of the earliest CNN-based approaches, utilizing three decoder branches with a shared encoder to perform nuclei separation, detection, and classification in H&E WSIs. The post-processing step of the three output branches generates the final detected and classified nuclei. TSFD-Net, a tissue-specific feature distillation pipeline for nuclei detection and classification, was proposed in [25]. TSFD-Net incorporated a tissue-specific feature distillation backbone together with a bidirectional feature pyramid network and interlinked decoders for hierarchical feature fusion and joint optimization. The framework predicts nuclei classification and boundary masks, followed by post-processing to generate the final output. Cerberus [23] is a multi-task learning framework with a shared encoder and task-specific decoders, enabling the prediction of nuclei, glands, lumen, and tissue regions in H&E WSIs. The architecture consists of four dedicated decoder branches, where each decoder is optimized for a specific task while sharing a common encoder representation for efficient feature learning and joint optimization. CellViT [18] introduced a ViT-based encoder architecture in which the backbone was integrated with three decoder branches through skip connections to jointly learn nuclei separation, detection, and classification. The outputs from these branches were fused through a post-processing stage to obtain the final output. Inspired by CellViT, CellViT++ [19] utilized an FM-based encoder to generate nuclei detection masks and deep nuclei feature embeddings, which were subsequently used by a lightweight classifier for nuclei classification. Similar to CellViT, NuLite [29] employed a lightweight U-Net-based encoder–decoder to reduce computational cost and inference time. The architecture utilized three decoder branches dedicated to nuclei detection, classification, and separation. LKCell [20] also utilized a U-Net-based encoder–decoder architecture with large-kernel convolutions and three segmentation heads dedicated to nuclei detection, classification, and separation. PointNu-Net [24] utilizes a high-resolution network backbone followed by three task-specific branches: a heatmap branch, a kernel branch, and a feature branch. These branches were integrated during post-processing to generate the final detected and classified nuclei. CellSAM [17] is an FM-based pipeline built on top of the Segment Anything Model [30]. It combines an automatic cell detector, CellFinder, with SAM to detect multiple cell types, organisms, and microscopy modalities using a single unified model. However, the framework focuses on cell detection, without performing cell classification. CellDETR [21] uses a hierarchical Swin transformer backbone to extract multi-scale feature representations. The extracted features are flattened and enriched with positional and level embeddings before being processed by a transformer encoder using multi-scale attention layers, detecting and classifying nuclei.

More details about the state-of-the-art pipelines are provided in the supplementary material (Table S2). Despite recent advances, several limitations remain in existing nuclei detection and classification pipelines. Many pipelines are trained and evaluated using intra-dataset settings, using train/test splits or cross-validation. Consequently, their cross-dataset generalizability remains unclear. In addition, benchmarking different pipelines remains difficult due to variations in datasets and predicted nuclei classes.

## III. Data Material

In this work, several H&E-stained datasets were used, namely CoNSeP [22], Lizard [31], NuCLS [32], PanopTILs [33], SegPath [34], PanNuke [35], and TNMI [36]. In addition, the ILCD dataset is proposed as part of this work. A detailed description of each dataset is provided in the Supplementary Material. These datasets differ in terms of organ, scanner, magnifications, annotation procedure, cell categories, and tile sizes. The SegPath dataset does not provide cell-level annotations; therefore, additional preprocessing procedures were applied to generate cell-level annotations and standardize all image tiles to a size of 256 × 256 pixels across datasets. Furthermore, all datasets were unified into three common cell categories by merging the original labels into immune cells, tumor cells, and other cells. This allows a fair and consistent evaluation of the different pipelines across multiple datasets.

## IV. Method

The overview of the proposed pipeline is presented in **Fig. 2**. Given an input H&E image tile, the hematoxylin (H) component is first separated and processed with a multi-scale Difference of Gaussians (DoG) filter to identify nuclei regions and generate a DoG mask. The DoG mask is then concatenated with the H&E image to obtain a 4-channel input, which is fed into the model. The model produces both a binary nuclei segmentation mask and a nuclei type classification mask. Finally, a post-processing step is applied to refine the boundaries of nuclei and generate the final predictions, where nuclei are detected and classified into tumor, immune, and other types. The main step of the pipeline is described below.

### *A. H Component Separation*

Given an H&E image $I \in R^{H \times W \times 3}$, the image is first converted into optical density space [37]:

$$O = -log\left(\frac{I+1}{I_o}\right), \tag{1}$$

where $I_o = 240$ is the transmitted light intensity, and the constant 1 avoids numerical instability from zero-pixel values. Pixels with low optical density values are removed using a threshold $\beta = 0.15$:

$$\hat{O} = O\,[O \geq \beta]. \tag{2}$$

The covariance matrix of the optical density matrix is computed, and then eigenvalue decomposition:

$$\Sigma = \mathrm{Cov}(\hat{O}^{\mathrm{T}}), \tag{3}$$

$$\Sigma\, v = \lambda . v, \tag{4}$$

where $\Sigma$ denotes the covariance matrix, $v$ is an eigenvector, and $\lambda$ is an eigenvalue. The $\hat{O}$ values are then projected onto this two-dimensional stain plane:

$$\Theta = \hat{O} . v . \tag{5}$$

The angular distribution of the projected pixels is computed:

$$\varphi = arctan2(\Theta) . \tag{6}$$

The minimum and maximum stain vectors are estimated using the 1 and 99 percentiles of the angular distribution.

$$\varphi_{min} = \mathrm{percentile}(\varphi, 1), \tag{7}$$

$$\varphi_{max} = \mathrm{percentile}(\varphi, 99). \tag{8}$$

The two stain vectors are computed:

$$v_{min} = v\,[cos(\varphi_{min}), sin(\varphi_{min})] \tag{9}$$

$$v_{max} = v\,[cos(\varphi_{max}), sin(\varphi_{max})] \tag{10}$$

The final stain matrix $M_{HE}$ is computed as:

$$M_{HE} = [v_{min} \quad , \quad v_{max}] . \tag{11}$$

The optical density matrix $O$ is then decomposed into stain concentrations using least-squares estimation:

$$\mathrm{C} = \underset{C}{\mathrm{argmin}} \,||\, O^T - M_{HE} . C||^2. \tag{12}$$

The first row of C, denoted as $C_H$, corresponds to the hematoxylin concentration. Then, the hematoxylin concentration is normalized using the 99$^{th}$ percentile:

$$C_{H,max} = \text{percentile}(C_H, 99)\ , \quad (13) \qquad C_{H,norm} = \frac{C_H}{C_{H,max}/C_{H,ref}}\ , \quad (14)$$

where $C_{H,ref} = 1.9705$ is the reference maximum hematoxylin concentration. Finally, the hematoxylin image is reconstructed using the hematoxylin reference stain vector:

$$H = I_o \exp(-H_{ref} \, . \, C_{H,norm})\ , \quad (15)$$

where $H_{ref}$ is the hematoxylin column of the reference stain matrix $[0.5626, 0.7201, 0.4062]^T$. The reconstructed $H$ image (**Fig. 2**) is then reshaped back to the original image size ($H$×$W$×3).

### *B. Difference of Gaussian (DoG)*

Given the extracted H component, the next step is to convert the H image to grayscale $H_g \in R^{H\times W}$. Gaussian filtering is then defined as:

$$L_\sigma(H) = G_\sigma * H_g \quad (16) \qquad G_\sigma(n,m) = \frac{1}{2\pi\,\sigma^2}\exp(-\frac{n^2-m^2}{2\,\sigma^2})\ , \quad (17)$$

where * denotes 2D convolution, $\sigma$ controls the smoothing scale, and $(n, m)$ denotes the spatial coordinates on $H_g$. For each Gaussian scale $\sigma$, the DoG response is computed by subtracting two Gaussian-smoothed versions of the $H_g$ image:

$$D_\sigma(H) = L_\sigma(H) - L_{k\sigma}(H)\ , \quad (18)$$

where $k = \sqrt{2}$ is the scale ratio between the two Gaussian kernels. DoG response is computed over a predefined set of Gaussian scales $S$. The final DoG mask is obtained by selecting the maximum response across all scales:

$$D(H) = \max_{\sigma \in S} D_\sigma(H),\ where\ S = \begin{cases} \{2.5, 3.5, 4.5\} \text{ for } 40x \\ \{1.6, 2.4, 3.2\} \text{ for } 20x \end{cases}, \quad (19)$$

where the scale set $S$ is defined according to the image magnification and obtained through several experiments for both 20× and 40× magnifications.

### *C. Model Architecture*

As shown in **Fig. 2**, the four-channel input is fed to a UniRepLKNet-N encoder [38], which extracts hierarchical feature representations. The encoder consists of four stages. Each stage progressively downsamples the spatial resolution while increasing the feature dimensions through large-kernel depth-wise convolutions. The decoder follows a U-Net-like architecture with skip connections between the encoder and decoder stages. High-level encoder features are progressively upsampled and fused with lower-level encoder features using concatenation blocks. Finally, two independent segmentation heads are attached to the decoder output. The first head predicts the binary nuclei segmentation mask, while the second head predicts the nuclei type classification mask. Following [18], the overall objective function is defined as:

$$L_{total} = L_{type} + L_{det}\ , \quad (20)$$

where $L_{type}$ and $L_{det}$ denote the nuclei type classification and nuclei detection losses, respectively. The nuclei type classification and nuclei detection losses are defined as:

$$L_{type} = \lambda_{CE}^{type} L_{CE}^{type} + \lambda_{Dice}^{type} L_{Dice}^{type} + \lambda_{FT}^{type} L_{FT}^{type}, \quad (21)$$

$$L_{det} = \lambda_{Dice}^{det} L_{Dice}^{det} + \lambda_{FT}^{det} L_{FT}^{det}, \quad (22)$$

where $L_{CE}^{type}$, $L_{Dice}^{type}$, and $L_{FT}^{type}$ denote the cross-entropy, Dice, and Focal Tversky losses for nuclei type classification, respectively, while $L_{Dice}^{det}$ and $L_{FT}^{det}$ denote the Dice and Focal Tversky losses for nuclei detection. The weighting parameters $\lambda$ control the contribution of each loss component during training and were set similarly to [18]. The Adam optimizer was employed to optimize the proposed loss function and train the network parameters. The model was trained for 150 epochs using a batch size of 24 and an initial learning rate of $1 \times 10^{-3}$. To improve model robustness and generalization, several data augmentation techniques were applied during training, including random horizontal and vertical flipping, random 90° rotations, color jittering, and Gaussian blurring. Training was performed on an Ubuntu 20.04 operating system, a 40-core CPU, and a 512 GB of RAM workstation equipped with an NVIDIA A100 GPU with 80 GB of memory running CUDA 12.2. To maintain consistency with existing pipelines, which are commonly trained on the multi-organ PanNuke dataset [35], CP-Net was also trained on the PanNuke dataset and further evaluated externally on the remaining datasets as detailed in section V.

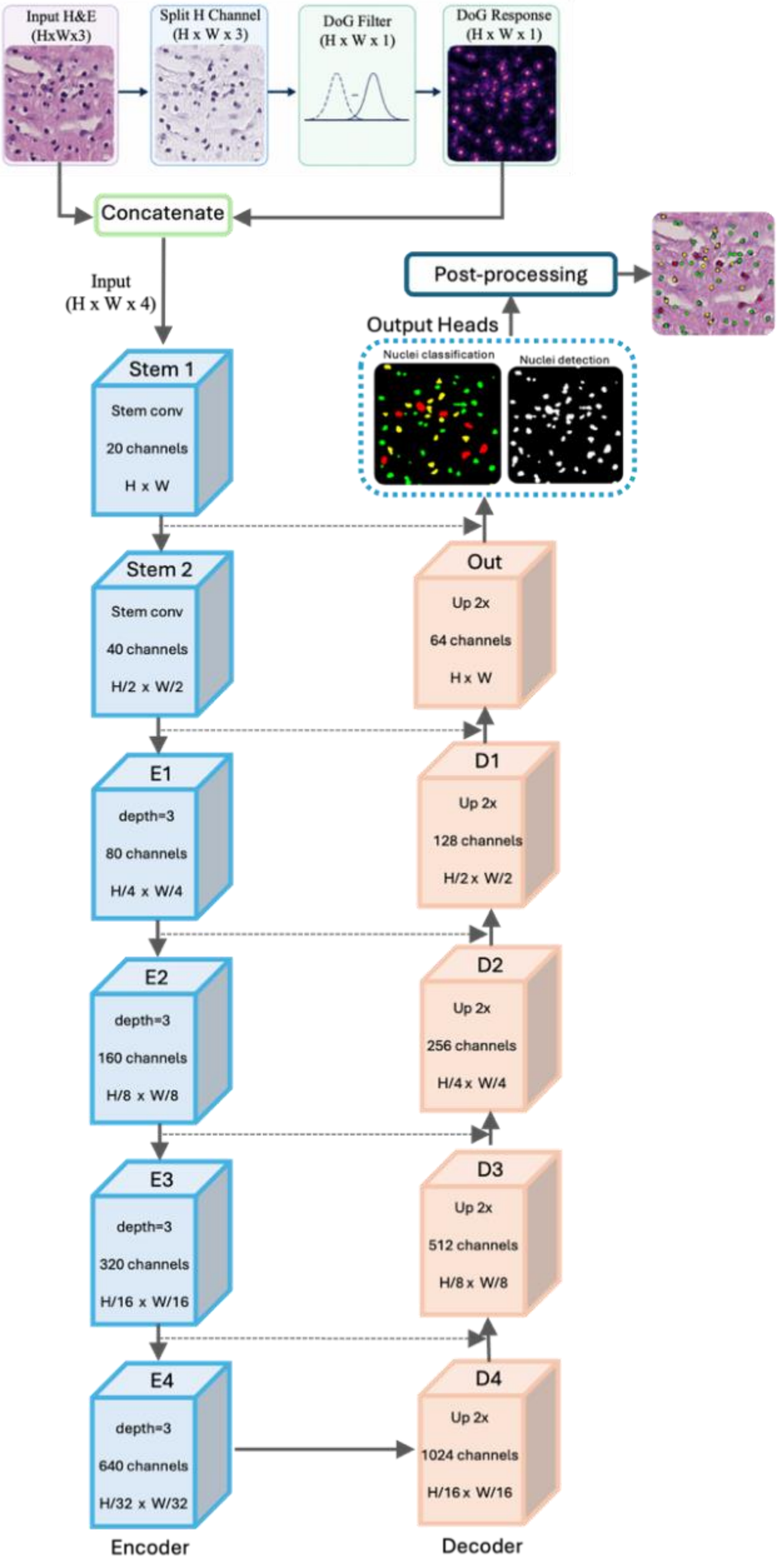

**Fig. 2**. Overview of the CP-Net architecture.

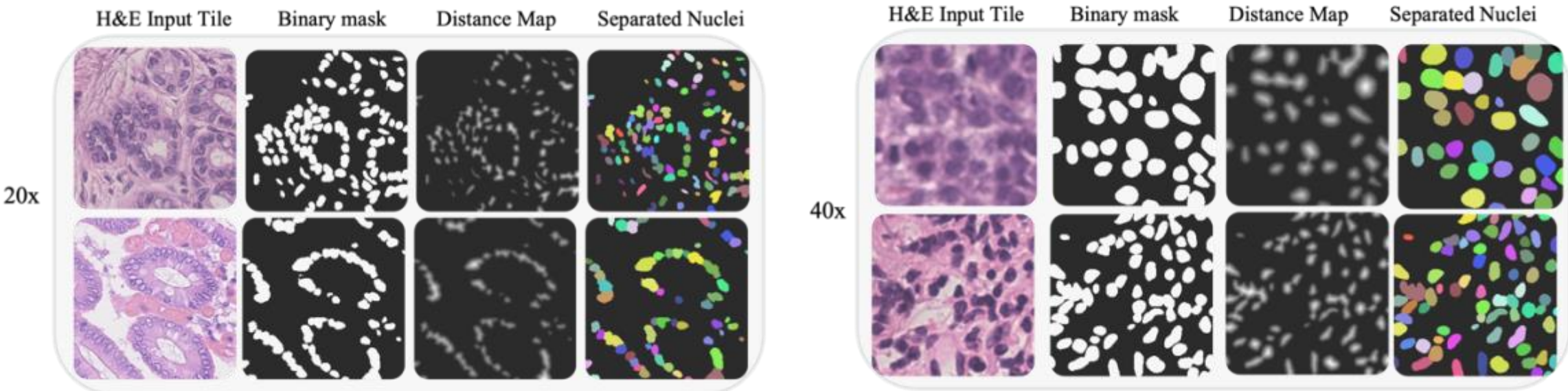

**Fig. 3**. Examples of the nuclei separation procedure (Steps 1–5) for 20× and 40× magnifications. From left to right: input H&E image tile, binary mask, distance mask, and final separated nuclei, where each nucleus is represented by a different color.

### D. Post-Processing

Consider the predicted nuclei mask $B$ for a given H&E image tile $I$, where 1: nuclei pixels and 0: background pixels. The following procedure was developed to separate touching nuclei and generate the final nuclei segmentation mask:

1) **Morphological erosion**: Apply the morphological erosion to separate weakly connected nuclei and reduce boundary noise:

$$B_e = B \ominus \tilde{S}, \quad (23)$$

where $B_e$ is the eroded binary mask, $\ominus$ denotes morphological erosion, $\tilde{S}$ and is a structuring element defined by the erosion radius $e_r$.

2) **Distance mask:** Compute the distance mask to identify the central regions of nuclei:

$$D_m = E(B_e), \quad (24)$$

where $E(.)$ denotes the Euclidean distance transform function applied to the eroded binary mask $B_e$ with a mask size $m_s$. Pixels near nuclear centers have larger values than pixels near nuclear boundaries or background regions (**Fig. 3**).

3) **Local markers:** Extract local markers from the distance mask to initialize watershed segmentation:

$$M = \hat{L}(D_m), \quad (25)$$

where $M$ represents the set of marker locations and $\hat{L}(.)$ denotes the peak-local-maxima function, where each detected maximum likely corresponds to the center of a nucleus. The parameter $d_m$ controls the minimum distance between neighboring peaks and $t_{abs}$ filters weak peaks.

4) **Watershed segmentation:** The watershed algorithm was applied to the distance mask to generate the segmentation mask with separated nuclei:

$$I_w = W(-D_m, M, B_e), \quad (26)$$

where $W(.)$ denotes the watershed function, and $I_w$ represents the final segmentation mask in which each nucleus is assigned a unique label (**Fig. 3**).

5) **Small object removal:** Small segmented objects with an area smaller than the predefined threshold $a_{min}$ were removed, as they likely correspond to background noise.

6) **Final segmentation mask**: The final segmentation mask is generated by combining the segmentation mask $I_w$ with the predicted nuclei type mask, where each segmented nucleus is assigned to its corresponding class (**Fig. 2**).

This procedure has several hyperparameters that must be carefully chosen. Therefore, a parameter optimization procedure was adopted to determine the optimal set of hyperparameters for each magnification level 20× and 40×. First, 100 binary segmentation masks were carefully selected for each 20× and 40× magnification (a total of 200 masks). The selected masks included diverse nucleus types, densities, and touching-nuclei scenarios from different organs and datasets. Each segmentation mask was manually inspected, and the nuclei were visually counted to obtain the ground-truth number of nuclei, including both touching and non-touching cells. The procedure steps 1-5 were then applied using different hyperparameter combinations, and the following optimization loss function was computed:

$$\mathcal{L} = \frac{1}{N}\sum_{i=1}^{N} |C_i^{gt} - C_i^{pr}|, \quad (27)$$

where $C_i^{gt}$ is ground-truth nuclei count, $C_i^{pr}$ represents predicted number of nuclei after step 5, $N$ is the total number of evaluated masks. The optimized hyperparameters for the 20× setting were $e_r = 1$, $m_s = 5 \times 5$, $d_m = 4$, $t_{abs} = 2$, and $a_{min} = 10$, and for the 40× setting, $e_r = 2$, $m_s = 5 \times 5$, $d_m = 8$, $t_{abs} = 2$, and $a_{min} = 20$. For efficiency, the post-processing procedure was implemented using GPU-based morphological operations and connected component analysis with PyTorch/CuPy, together with multi-threaded watershed segmentation. **Fig. 3** presents representative nuclei separation results at both 20× and 40× magnifications, demonstrating that the proposed post-processing procedure effectively separates touching nuclei regardless of the magnification on which the model was trained.

## V. Performance Evaluation

The performance of the CP-Net was evaluated and compared with state-of-the-art methods, including CellViT, CellViT++, Cerberus, HoVer-Net, NuLite, PointNu-Net, CellDetr, CellSAM, LKCell, and TSFD (Table S2 in the Supplementary Materials). The evaluation was conducted on 8 datasets, including CoNSeP, ILCD, Lizard, NuCLS, PanopTILs, Segpath,

TNMI-20×, and TNMI-40× (Table S1 in the Supplementary Materials). Since different pipelines may predict different nuclei categories, the output classes of all pipelines were merged into three classes: immune cells, tumor cells, and other cells, to allow a consistent comparison. The performance metrics Detection Quality (DQ), Segmentation Quality (SQ), and Panoptic Quality (PQ) were used for evaluation [18][19][22]. DQ measures the accuracy of nuclei detection, SQ measures the quality of the predicted nuclei segmentation, and PQ is a unified metric that jointly evaluates nuclei detection and segmentation performance, reflecting how well the model detects and classifies nuclei. These metrics were computed separately for each nuclei class and then averaged across all classes.

**Fig. *4*** presents a comprehensive benchmarking of state-of-the-art pipelines across multiple datasets (a), scanner types (b), nuclei classes (c), magnifications (d), performance vs inference time (e) averaged across all datasets. A qualitative comparison of each dataset and pipeline is provided in Table S3 of the Supplementary Materials. The following conclusion can be drawn:

- **Across datasets**: The DQ, SQ, and PQ metrics for each pipeline and dataset are presented in **Fig. *4*a**: colon (CoNSeP and Lizard), liver (ILCD), breast (NuCLS and PanopTILs), lung (TNMI), and the multi-organ (SegPath). The average performance across datasets ranged from 0.07–0.46 for DQ, 0.22–0.62 for SQ, and 0.05–0.34 for PQ, highlighting the large variability in metric values and the limited generalizability of current pipelines across different datasets. Pipelines trained on one dataset often struggle to generalize to unseen datasets because datasets differ in organ, staining, scanner, and tissue [22][17][21]. CP-Net, CellViT, and NuLite achieved the best overall performance.
- **Scanner type**: The average PQ metric for each pipeline across different scanner types is presented in **Fig. *4*b**, including Omnyx VL120 (CoNSeP), Hamamatsu Nanozoomer S60 (SegPath), 3DHistech Pannoramic Flash III (TNMI), and unknown scanner types (ILCD, Lizard, NuCLS, and PanopTILs). The variability in performance across scanner types highlights the impact of scanner type on performance. Most pipelines achieved better performance on the Omnyx VL120 scanner, which may be attributed to CoNSeP being a single-dataset benchmark, whereas the other scanner groups contain multiple datasets. Different scanners produce variations in color distribution, contrast, sharpness, noise, and artifacts, leading to domain shifts between datasets [22][21][25].
- **Nuclei type**: The average PQ metric for each pipeline across different nuclei type are presented in **Fig. *4*c**. Most pipelines achieved better performance for tumor type, followed by immune cells and then other cell types. Tumor nuclei are often easier to recognize because they are larger and morphologically distinctive, while immune and other cells are smaller, denser, and more heterogeneous [18][25].
- **Magnification:** The average PQ metric for each pipeline across different magnification levels is presented in **Fig. *4*d**. Most pipelines achieved better performance at 40× compared with 20× magnification. The nuclei at 40× appear larger and contain clearer morphological and texture information, with more visible boundaries. Also, most existing pipelines were originally developed and trained primarily on 40× datasets [22][18][20].
- **Performance *vs* efficiency**: **Fig. *4*e** presents the trade-off between inference time and PQ for the evaluated pipelines. CP-Net, CellDETR, and CellViT form the Pareto frontier. Among them, CP-Net achieved the most favorable balance, providing high PQ performance (slightly lower than CellViT) while maintaining low inference time compared with the other evaluated pipelines.

## VI. CellQuant-Net for Nuclei Quantification

**Fig. *5*** presents the proposed CellQuant-Net that integrates WSI quality assessment (QA), nuclei detection and classification, spatial neighborhood connectivity and quantification analysis. First, the QA pipeline [26] identifies high-quality tissue regions and excludes artifacts such as irrelevant background, blurred, folded, and pen-marked regions. Next, CP-Net is applied to detect and classify nuclei. Then, the neighborhood connectivity analysis [39] was applied to generate a spatial nuclei connectivity metric. Each nucleus is connected to its neighboring nuclei within a predefined radius. Finally, the extracted spatial connectivity information is used to compute several quantitative metrics, including cell counts, percentages, densities, and immune infiltration score. In addition, CP-Net was trained on individual datasets to provide single-organ and multi-organ models for downstream computational pathology analysis and AI-assisted clinical interpretation. The pipeline and pretrained model are publicly available on GitHub.

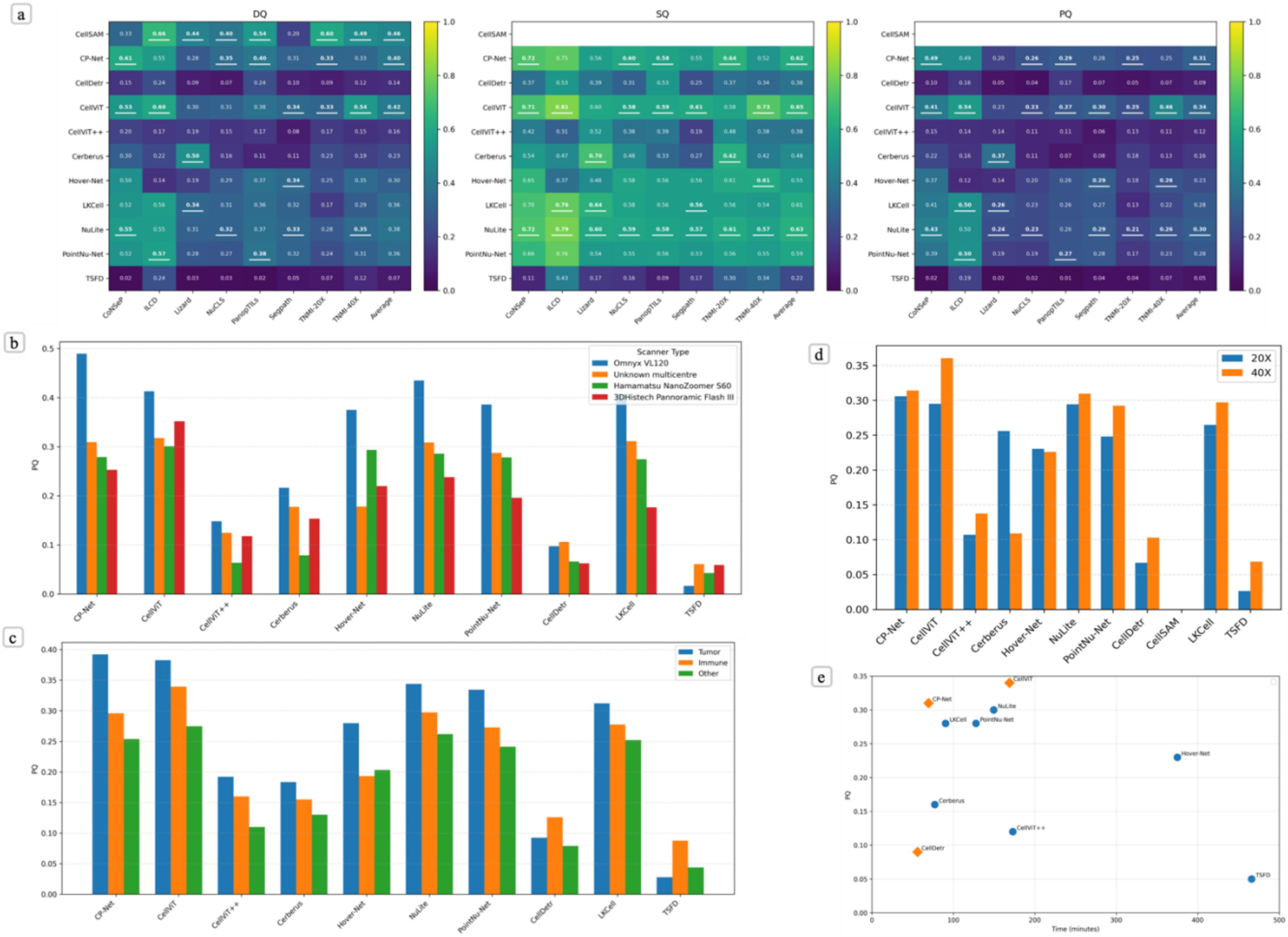

**Fig. 4**. Comprehensive benchmarking of nuclei detection and classification pipelines across multiple datasets, scanner types, magnifications, and nuclei types. (a) DQ, SQ, and PQ for each pipeline the evaluated datasets, with the average performance shown in the final column. Since CellSAM only performs nuclei detection, only DQ results are available. For each dataset, the top three performing pipelines are highlighted (underlined in white). (b) Average PQ performance grouped by scanner type. (c) Class-wise PQ comparison for tumor, immune, and other nuclei type. (d) PQ comparison between 20× and 40× magnifications. (e) Trade-off between inference time and PQ (averaged over datasets) for the evaluated pipelines. Pipelines marked by orange squares, CP-Net, CellDetr, and CellViT indicate the Pareto frontier. The best trade-off between performance and computational efficiency is obtained for CP-Net.

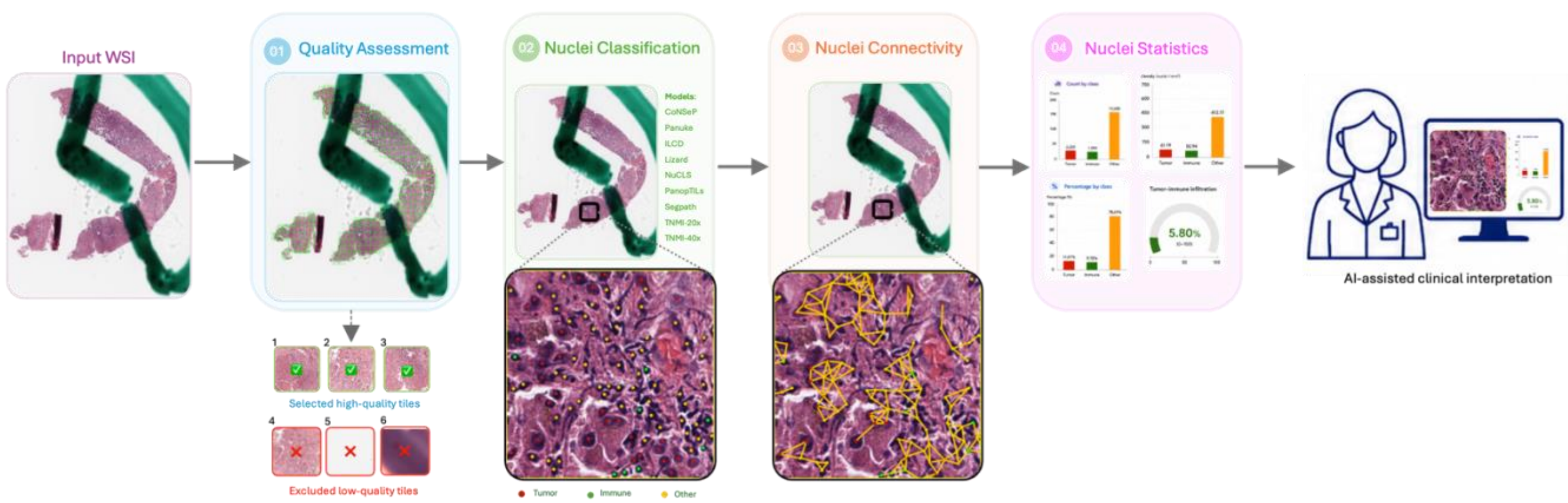

**Fig. 5**. Overview of the CellQuant-Net pipeline for nuclei quantification: (1) WSI quality assessment (QA) to identify high-quality tissue tiles {1-3} and exclude tiles with artifacts {blurred:4; background:5; fold:6}; (2) nuclei detection and classification using CP-Net to identify nuclei types; (3) spatial nuclei connectivity analysis to characterize cell-cell relationships; (4) statistical analysis to compute nuclei quantification metrics.

## VII. Conclusion And Future Work

This paper presents CellPrior-Net (CP-Net) for H&E-WSI nuclei detection and classification. By incorporating hematoxylin-based prior information, CP-Net guides the network toward nuclei-relevant regions while maintaining a lightweight architecture and fast inference speed. Extensive benchmarking on 8 public and private datasets obtained from different scanners, organs, magnifications, and clinical centers demonstrated that CP-Net achieves competitive performance compared with state-of-the-art pipelines while maintaining high computational efficiency. The results also indicate that organ type, scanner type, cell type, tissue composition, and magnification can substantially influence model performance.

In addition, CellQuant-Net is introduced, an end-to-end nuclei quantification pipeline that integrates WSI quality assessment, nuclei detection and classification, and spatial neighborhood analysis. The pipeline allows the extraction of several quantitative nuclei metrics, facilitating downstream computational pathology analysis.

Although CP-Net achieved competitive performance, several limitations and aspects for enhancement remain. Future work will therefore focus on the following directions:

- Collect additional multi-center datasets covering different organs, scanner types, staining procedures, magnifications, cancer grades, and histological subtypes to support the development, training, testing, and standardized cross-dataset benchmarking of different pipelines under a common evaluation framework, thereby improving model robustness and generalization.
- Extend benchmarking beyond detection and classification metrics by exploring the association between cell quantification results with independent biological measurements obtained from orthogonal modalities, including IHC and RNA expression data, to further evaluate the biological relevance and clinical utility of the proposed pipeline.
- Extend CP-Net to support a broader range of nuclei categories by leveraging immunofluorescence (IF) guided annotations and co-registered H&E–IF datasets, enabling more accurate cell-type labeling and improved nuclei detection and classification performance.
- Use advanced 3D pathology imaging to move beyond 2D analysis toward 3D nuclei detection and classification, enabling more accurate characterization of nuclei morphology, while providing richer features to further improve model performance.

### Ethics Statement

This study used internal datasets, for which all patients provided informed consent and data collection was approved by the Institutional Review Board at each participating institute (ILCD: R16008; TNM-I: TNM-I-REK2016/2054), as well as publicly available anonymized datasets. Therefore, no additional ethical approval or informed consent was required, and all data were used in accordance with the terms set by the original data providers.

### Declaration of Competing Interest

Dr. Pinato reported lecture fees from Bayer Healthcare, AstraZeneca, EISAI, Bristol Myers Squibb, Roche, Boston Scientific, and Ipsen; travel support from Bristol Myers Squibb, Roche, Bayer Healthcare, and AstraZeneca; and consulting fees from Accession Therapeutics, MAIA Therapeutics, Bristol Myers Squibb, Mina Therapeutics, Galapagos, Parabilis, Boehringer Ingelheim, Ewopharma, EISAI, Ipsen, Roche, H3 Biomedicine, AstraZeneca, DaVolterra, Mursla, Terumo, Avammune Therapeutics, Starpharma, LiFT Biosciences, and Exact Sciences. All outside the submitted work. Dr Rakaee reported honoraria from AstraZeneca. No other COIs were reported.

### Declaration of generative AI

During the preparation of the first draft of this manuscript, the corresponding author used ChatGPT-4o for grammar check and readability of the text. Afterwards, the manuscript was reviewed and edited by the co-authors as needed. The corresponding author then performed a final review and editing of the manuscript, taking full responsibility for the content of the published article. Generative AI was not used to write the content.

### Acknowledgements

This research is supported by the Northern Norway Regional Health Authority (grant HNF1771-26 and HNF1660-23). The funders had no role in the design and conduct of the study, nor in the decision to submit the manuscript for publication.

## Data/code availability

The annotated image tiles for the TNMI and ILCD datasets will be made available for research purposes upon request. The remaining datasets used in this study are publicly available: PanNuke, CoNSeP, Lizard, NuCLS, PanopTILs, and SegPath. The source code for the proposed pipeline is publicly available on GitHub.

## Supplementary Material

**ConSeP [1]**: This dataset consists of 41 image tiles collected from colon cancer patients at the University Hospitals Coventry and Warwickshire, UK. The original tile size is 1000 × 1000 pixels with a spatial resolution of 0.25 µm/px at 40× magnification. The annotations were obtained by a pathologist and subsequently reviewed by a second pathologist to achieve consensus. The dataset contains multiple nuclei categories, including inflammatory, epithelial, fibroblast, muscle, and endothelial cells.

**ILCD**: This dataset consists of 19,016 image tiles collected from 40 WSIs liver cancer patients at Imperial College London across multiple centers. Each tile has a size of 256 × 256 pixels with a spatial resolution of 0.25 µm/px at 40× magnification. The dataset contains three cell categories: immune cells, tumor cells, and other cells. The annotations were initially performed by two PhD students and one postdoctoral researcher and were subsequently reviewed by a pathologist to ensure annotation quality and accuracy.

**Lizard [2]**: This dataset extends the CoNSeP dataset by integrating colon cancer tissue images collected from multiple centers in the USA and four hospitals in China. The dataset has a spatial resolution of 0.50 µm/px using 20× magnification and contains 270 image tiles of varying sizes, with an average dimension of 1016 × 917 pixels. The annotations were generated through automatic, semi-automatic, and manual boundary refinement procedures. The dataset contains multiple nuclei categories, including neutrophils, lymphocytes, plasma cells, eosinophils, epithelial cells, and connective tissue cells.

**NuCLS [3]**: This dataset is a crowdsourced collection of nuclei annotations from breast cancer tissue obtained from TCGA, annotated by medical students and pathologists. The dataset was acquired at a spatial resolution of 0.20 µm/px using 20× magnification and contains 151 large image tiles of varying sizes, with an average dimension of 3300 × 2500 pixels. The dataset includes multiple tissue and cellular categories, including tumor, stroma, lymphocytic infiltrate, necrosis or debris, glandular secretions, blood, metaplasia NOS, fat, plasma cells, other immune infiltrate, mucoid material, normal acinus or duct, lymphatics, nerve, skin adnexa, blood vessel, angioinvasion, ductal carcinoma in situ (DCIS), and other tissue regions.

**PanopTILs [4]**: This dataset provides region and cell-level annotations obtained from breast cancer patients within the TCGA cohort. The dataset was acquired at a spatial resolution of 0.25 µm/px using 40× magnification and consists of 157 large image tiles, each 1024 × 1024 pixels. It includes pathologist-approved nuclei annotations for multiple cell types, including TILs, stromal cells, epithelial cells, miscellaneous cells, plasma, and other.

**SegPath [5]:** This dataset consists of 158,687 registered H&E and immunofluorescence (IF) image patches collected from 18 organs at the University of Tokyo Hospital. The images have a size of 984 × 984 pixels with a spatial resolution of 0.22 µm/px at 40× magnification. The annotations were automatically generated through IF-based registration rather than manual pathologist annotation. The dataset provides region-wise segmentation masks for epithelium (Pan-cytokeratin), smooth muscle/myofibroblast (αSMA), lymphocytes (CD3/CD20), leukocytes (CD45RB), blood/lymphatic vessels (ERG), plasma cells (MIST1), myeloid cells (MNDA), and red blood cells (CD235a).

**PanNuke [6]**: This dataset consists of 7904 image tiles collected from 19 different organs. The images have a size of 256 × 256 pixels with a spatial resolution of 0.25 µm/px at 40× magnification. The dataset contains annotated nuclei categorized into five major cell types, including neoplastic, inflammatory, connective tissue, dead, and epithelial cells.

**TNMI [7]**: This dataset consists of 31,742 image tiles at 20× magnification and 18,424 image tiles at 40× magnification collected from lung cancer patients enrolled in the Scandinavian multi-institutional TNM-I clinical trial. The images have a size of 256 × 256 pixels with a spatial resolution of 0.50 µm/px at 20× magnification and 0.25 µm/px at 40× magnification. The dataset contains annotated nuclei categorized into three major cell types, including tumor, immune, and other cells.

**Table S1**. Summary of the datasets used in this study, including the number of image tiles, distribution of annotated cell types (tumor, immune, and other), total number of annotated cells, image magnification, scanner type, and organ. The datasets cover both single-organ and multi-organ histopathology cohorts obtained using different scanners and magnification settings.

| Dataset | No. of Tiles | No. of Tumor | No. of Immune | No. of Other | Total No. of Cells | Magnification | Scanner Type | Organ |
|---|---|---|---|---|---|---|---|---|
| **ConSeP** | 983 | 14310 | 9291 | 16512 | 40113 | 40x | Omnyx VL120 | Colon |
| **ILCD** | 19016 | 120979 | 195921 | 39412 | 356312 | 40x | Unknown-multicenter | Liver |
| **Lizard** | 7429 | 389722 | 226541 | 178071 | 794334 | 20x | Unknown-multicenter | Colon |
| **NuCLS** | 30464 | 760245 | 356774 | 422780 | 1539799 | 20x | Unknown-multicenter | Breast |
| **PanopTILs** | 1743 | 22244 | 9519 | 20592 | 52355 | 40x | Unknown-multicenter | Breast |
| **SegPath** | 542238 | 1837707 | 934981 | 858400 | 3631088 | 40x | Hamamatsu Nanozoomer S60 | Multi-organ |
| **PanNuke** | 7904 | 85279 | 28555 | 49137 | 162971 | 40x | Unknown-multicenter | Multi-organ |
| **TNMI-20** | 31742 | 1974156 | 276707 | 1025019 | 3275882 | 20x | 3DHistech Pannoramic Flash III | Lung |
| **TNMI-40** | 18424 | 403213 | 25156 | 119071 | 547440 | 40x | 3DHistech Pannoramic Flash III | Lung |

**Table S2**. Comparison of state-of-the-art nuclei detection and classification pipelines in digital pathology, including architecture type, model size, and supported tasks.

| Pipeline | CP-Net | CellViT | CellViT++ | Cerberus | HoverNet | NuLite | PointNu-Net | CellDetr | CellSam | LKCell | TSFD |
|---|---|---|---|---|---|---|---|---|---|---|---|
| Parameters | 46.4 M | 699.7 M | 699.7 M | 28.4 M | 37.6 | 47.9 | 161.4 | 207.9 | 289.6 | 163.8 | 22.1 |
| Architecture | CNN | ViT | ViT | CNN | CNN | CNN | CNN | ViT | ViT | CNN | CNN |
| Task | Detection & classification | Detection & classification | Detection & classification | Detection & classification | Detection & classification | Detection & classification | Detection & classification | Detection & classification | Detection | Detection & classification | Detection & classification |

**Table S3**. Comparing the ground truth (GT) annotations with predictions generated by different state-of-the-art pipelines for each dataset.

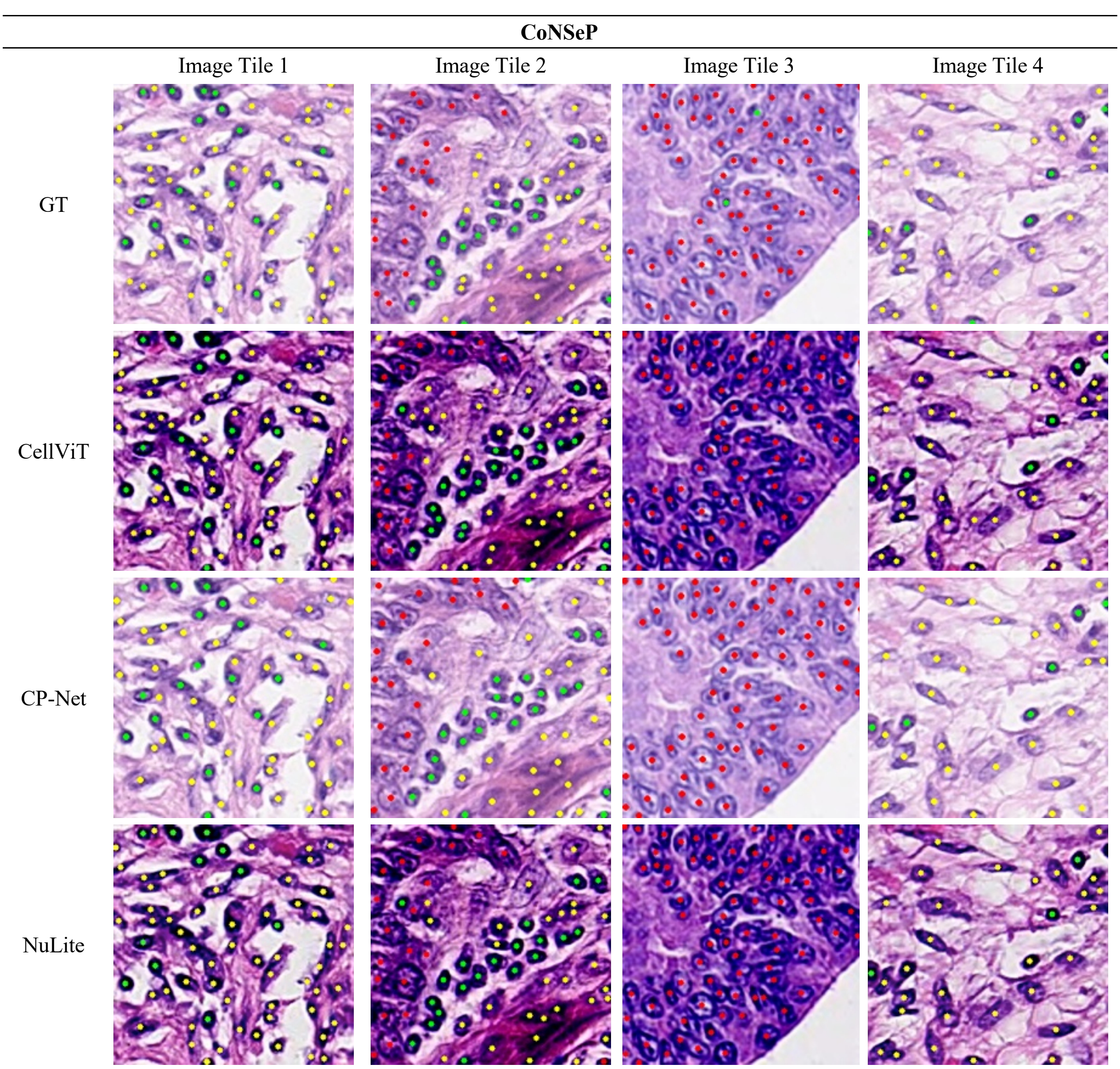

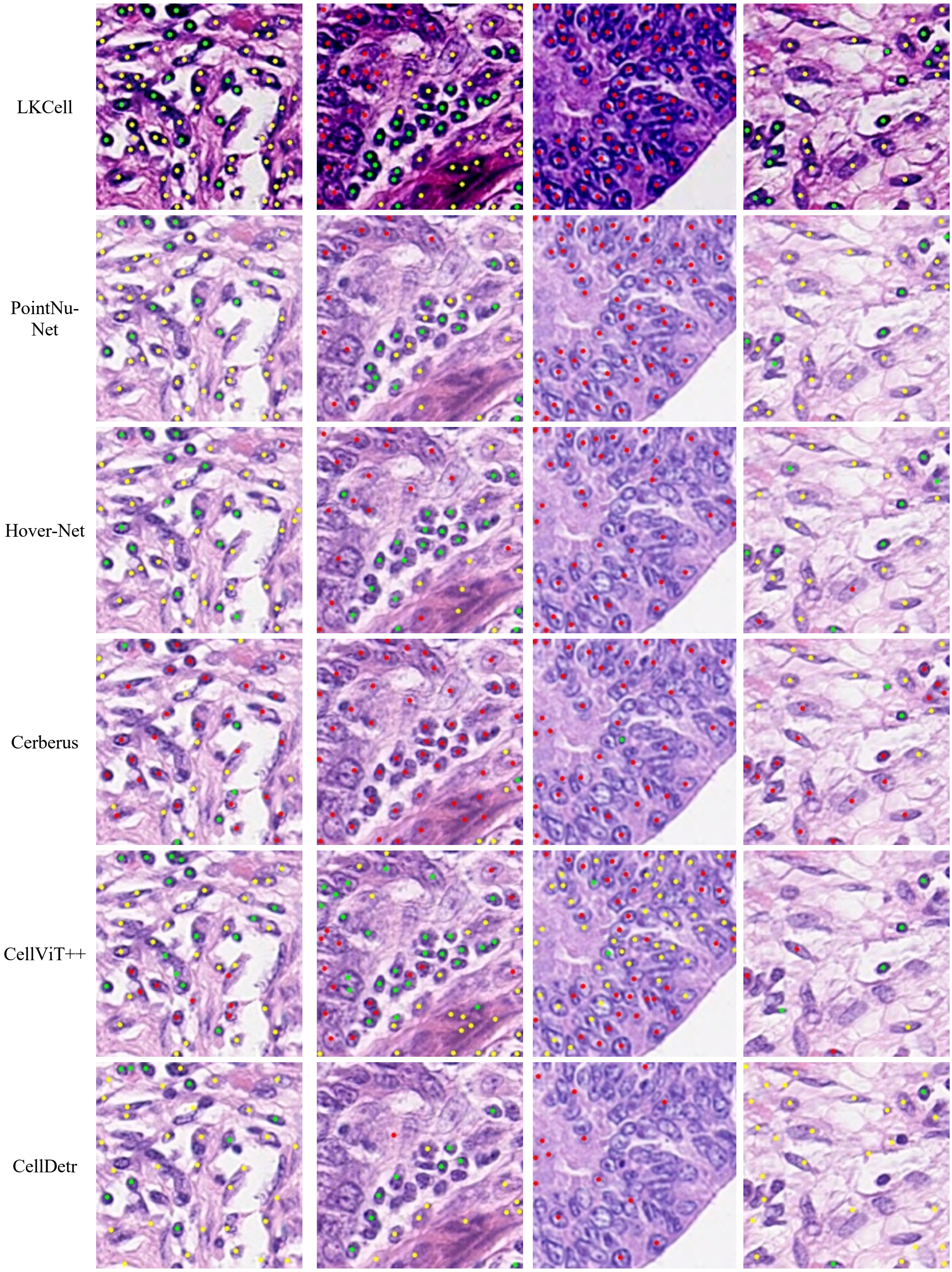

LKCell
PointNu-Net
Hover-Net
Cerberus
CellViT++
CellDetr

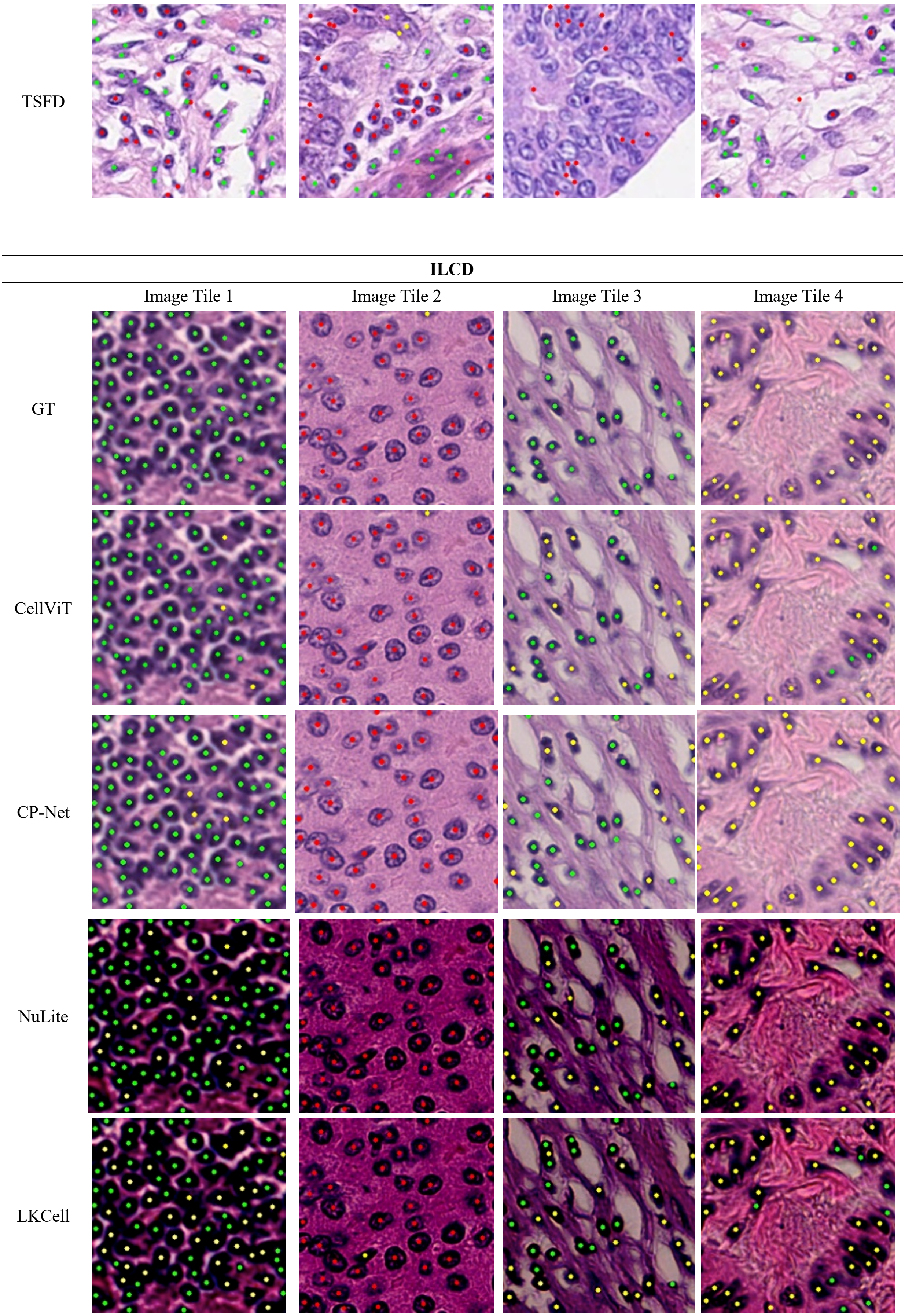

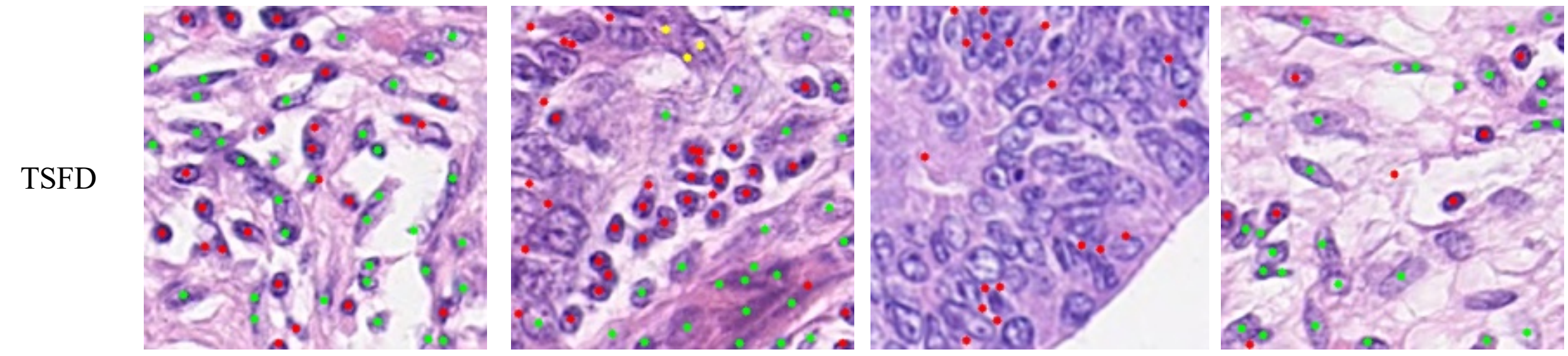

**ILCD**

| | Image Tile 1 | Image Tile 2 | Image Tile 3 | Image Tile 4 |
|---|---|---|---|---|
| GT | | | | |
| CellViT | | | | |
| CP-Net | | | | |
| NuLite | | | | |
| LKCell | | | | |

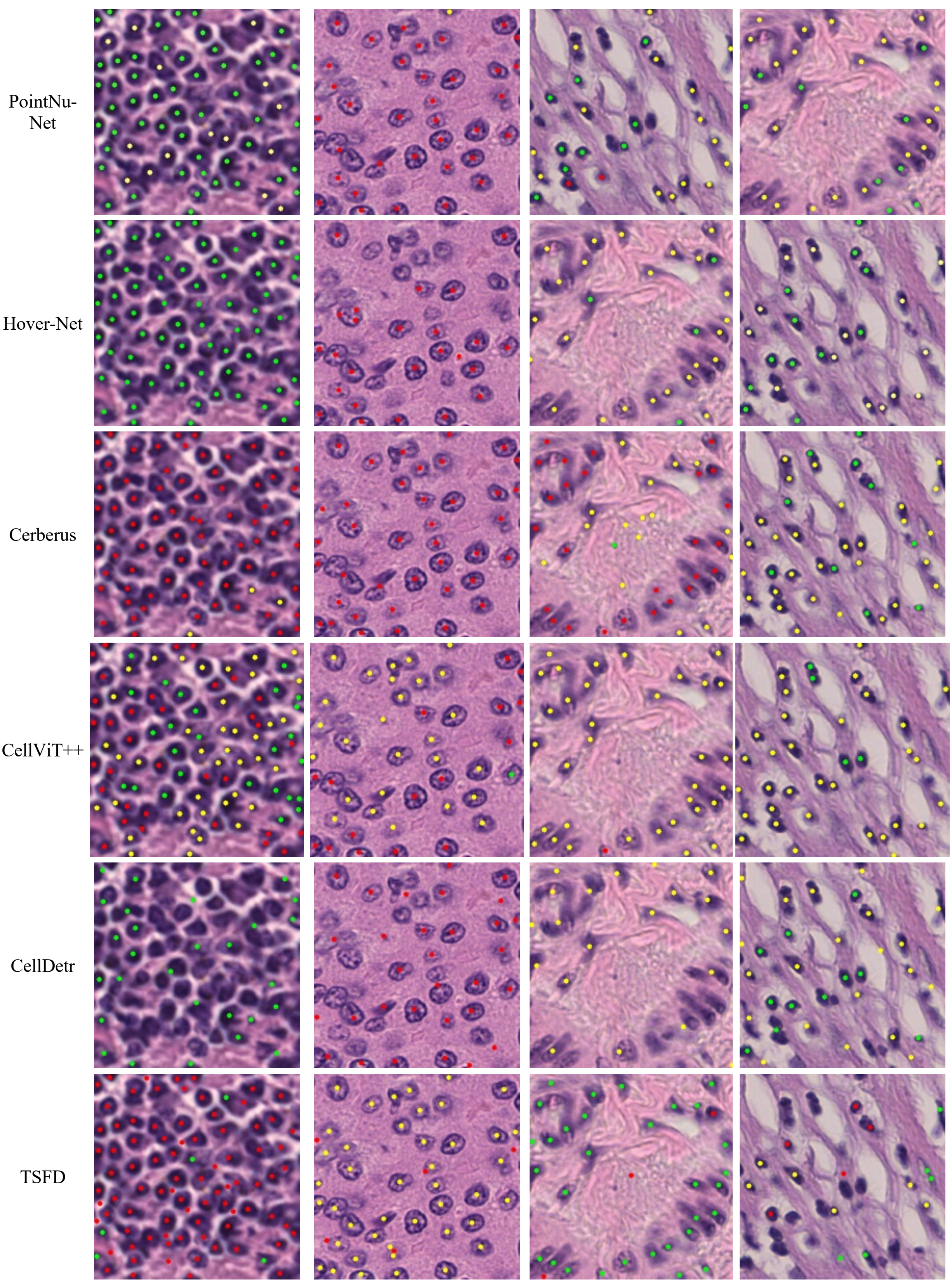

PointNu-Net
Hover-Net
Cerberus
CellViT++
CellDetr
TSFD

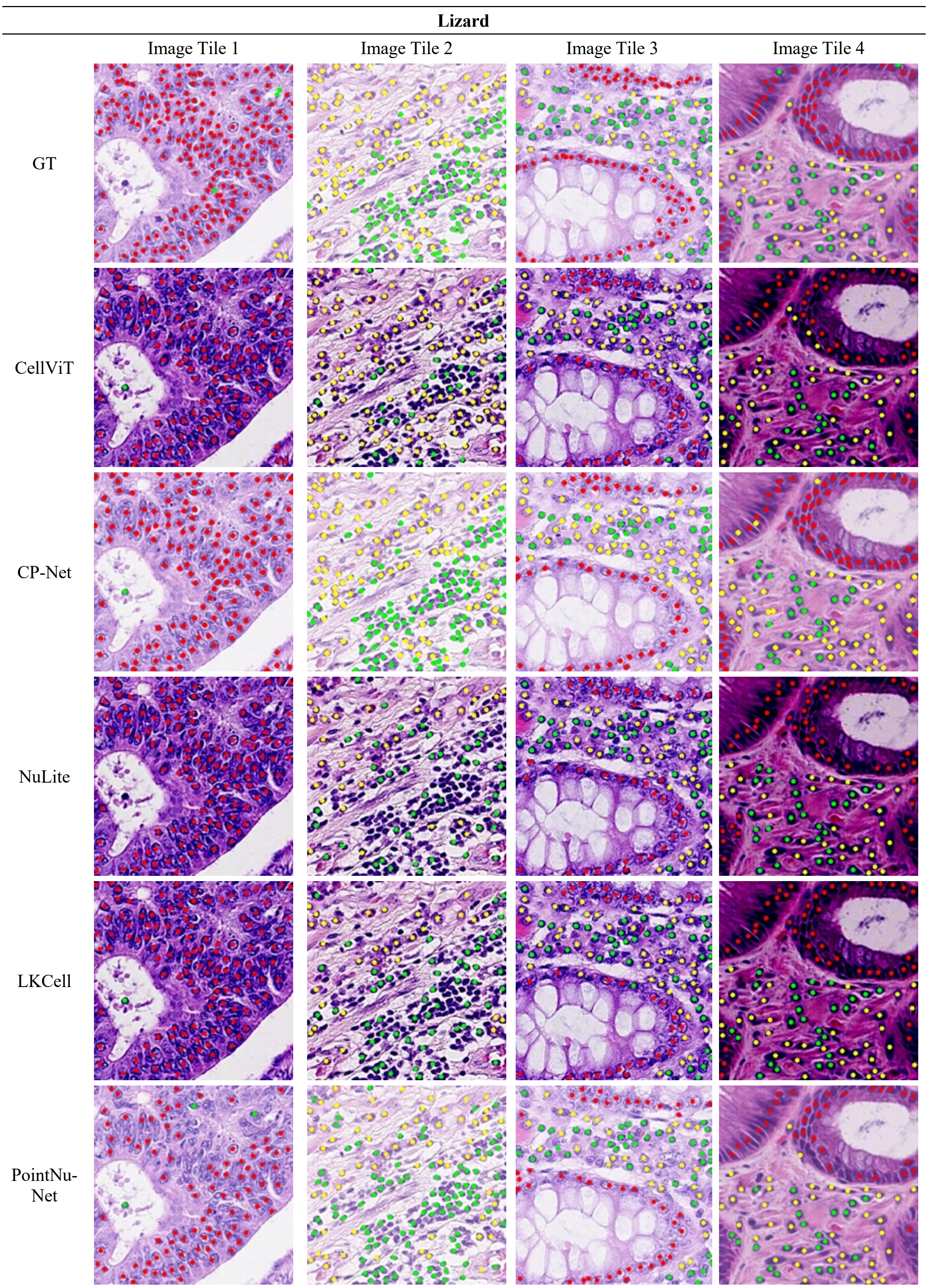
Lizard
Image Tile 1
Image Tile 2
Image Tile 3
Image Tile 4
GT
CellViT
CP-Net
NuLite
LKCell
PointNu-Net

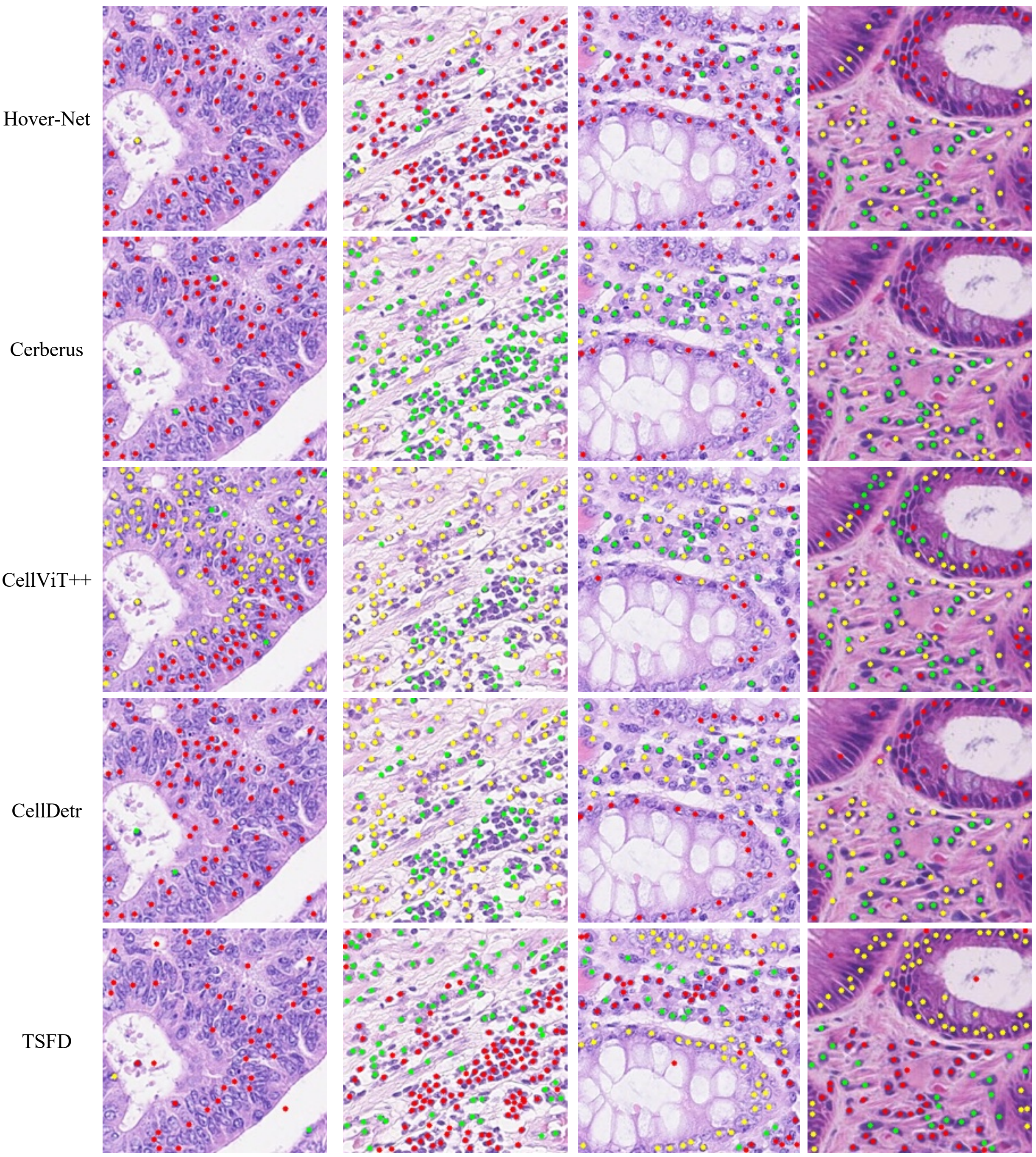
Hover-Net
Cerberus
CellViT++
CellDetr
TSFD

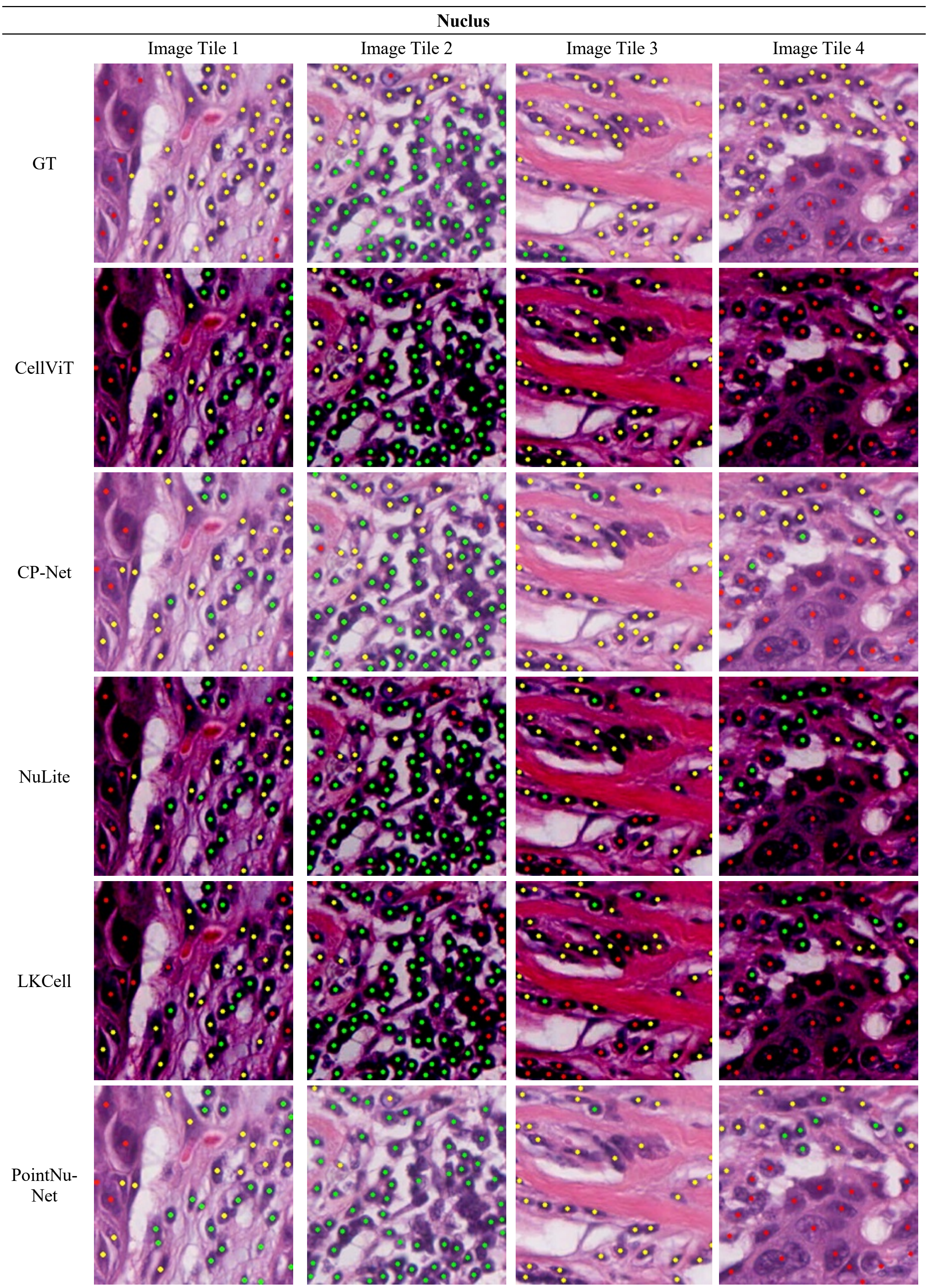
Nuclus
Image Tile 1
Image Tile 2
Image Tile 3
Image Tile 4
GT
CellViT
CP-Net
NuLite
LKCell
PointNu-Net

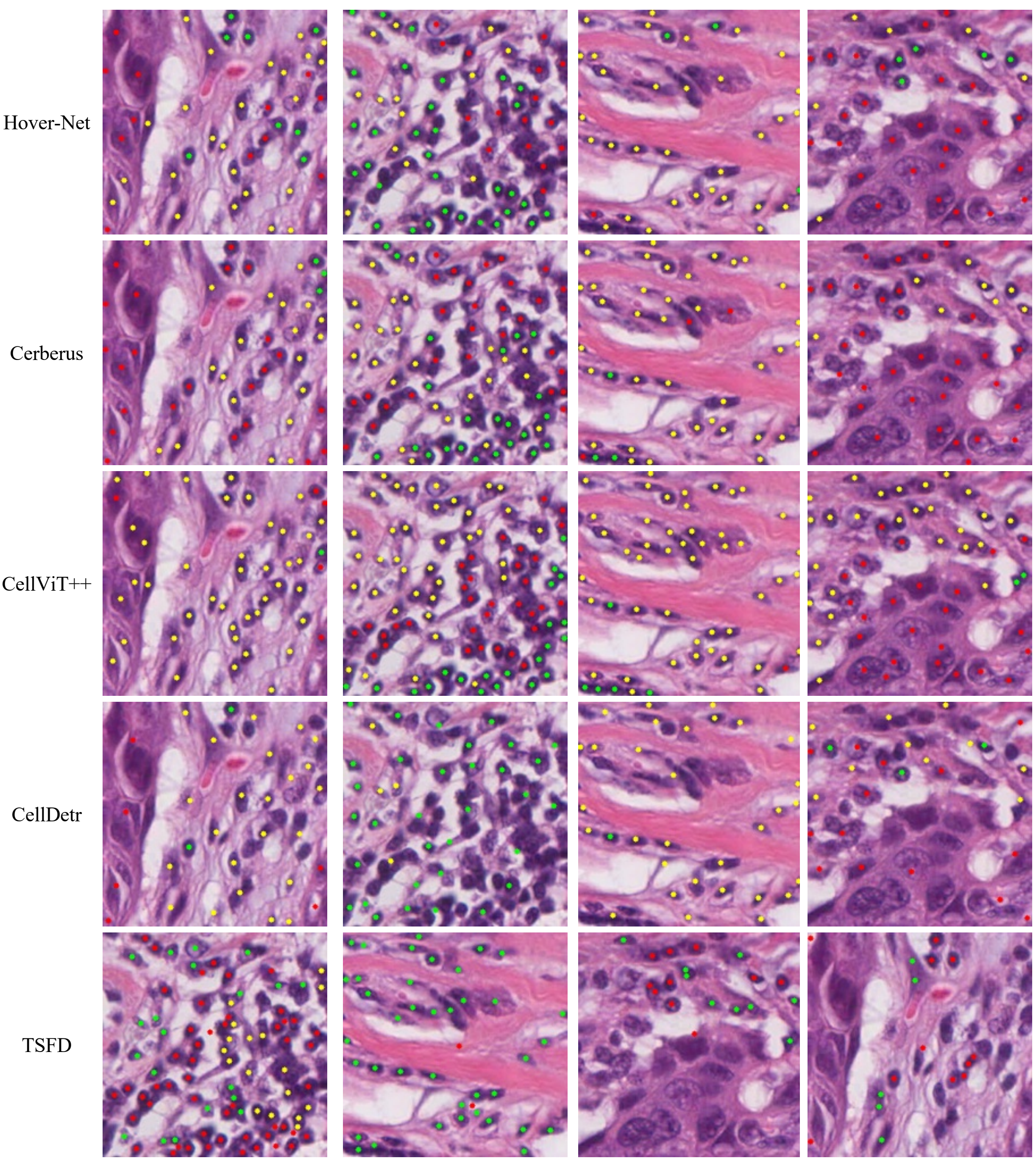
Hover-Net
Cerberus
CellViT++
CellDetr
TSFD

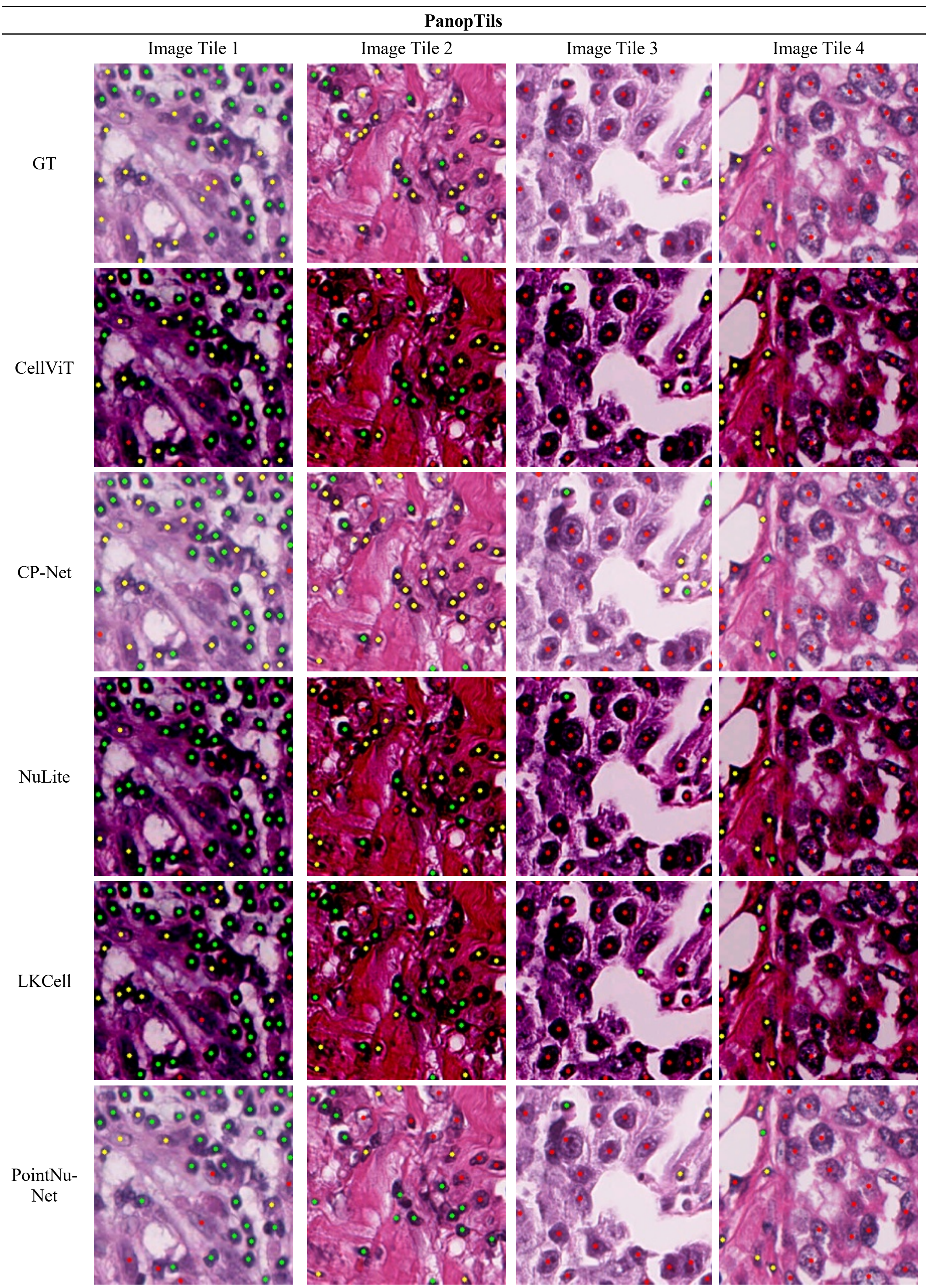
PanopTils
Image Tile 1
Image Tile 2
Image Tile 3
Image Tile 4
GT
CellViT
CP-Net
NuLite
LKCell
PointNu-Net

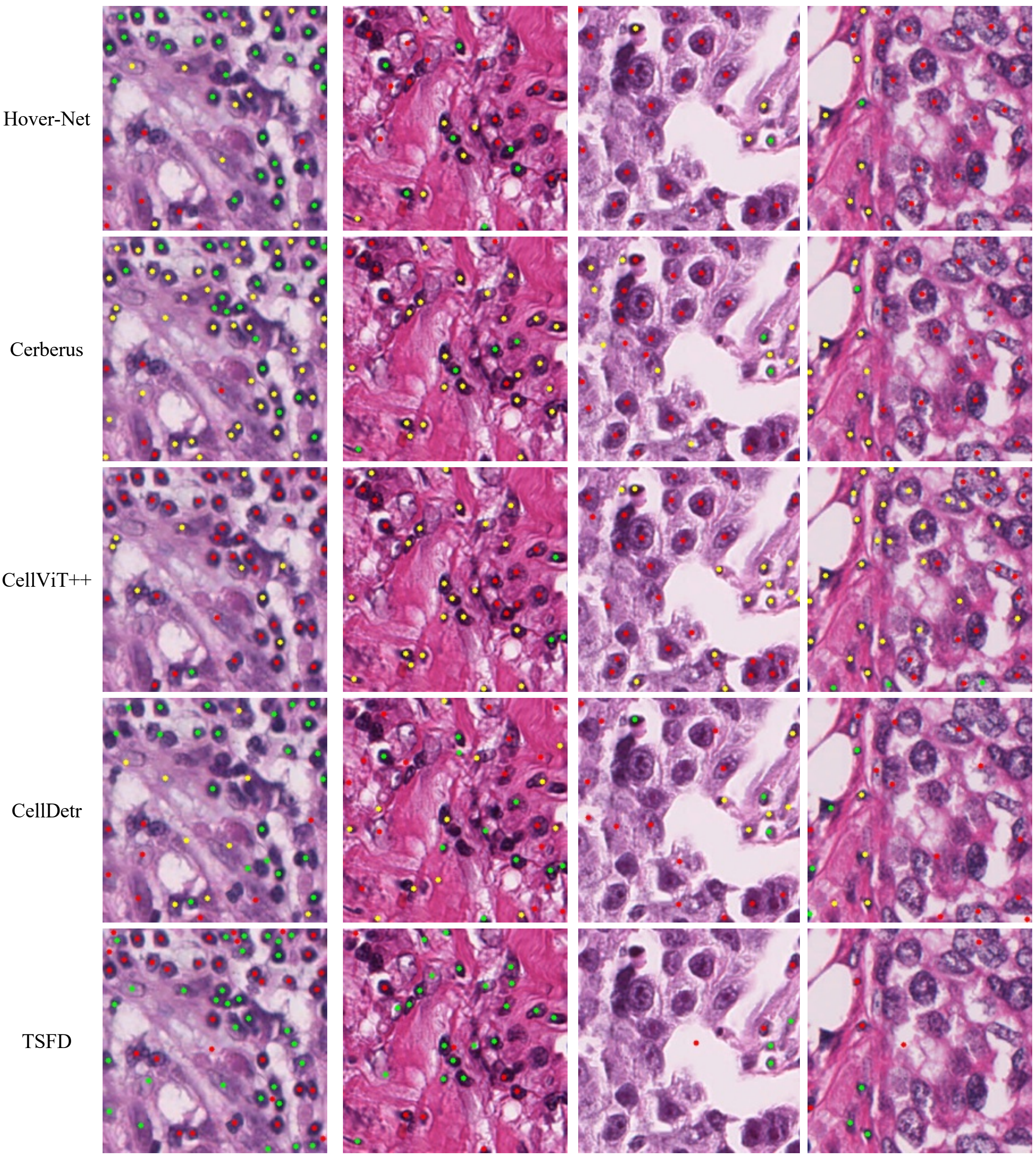
Hover-Net
Cerberus
CellViT++
CellDetr
TSFD

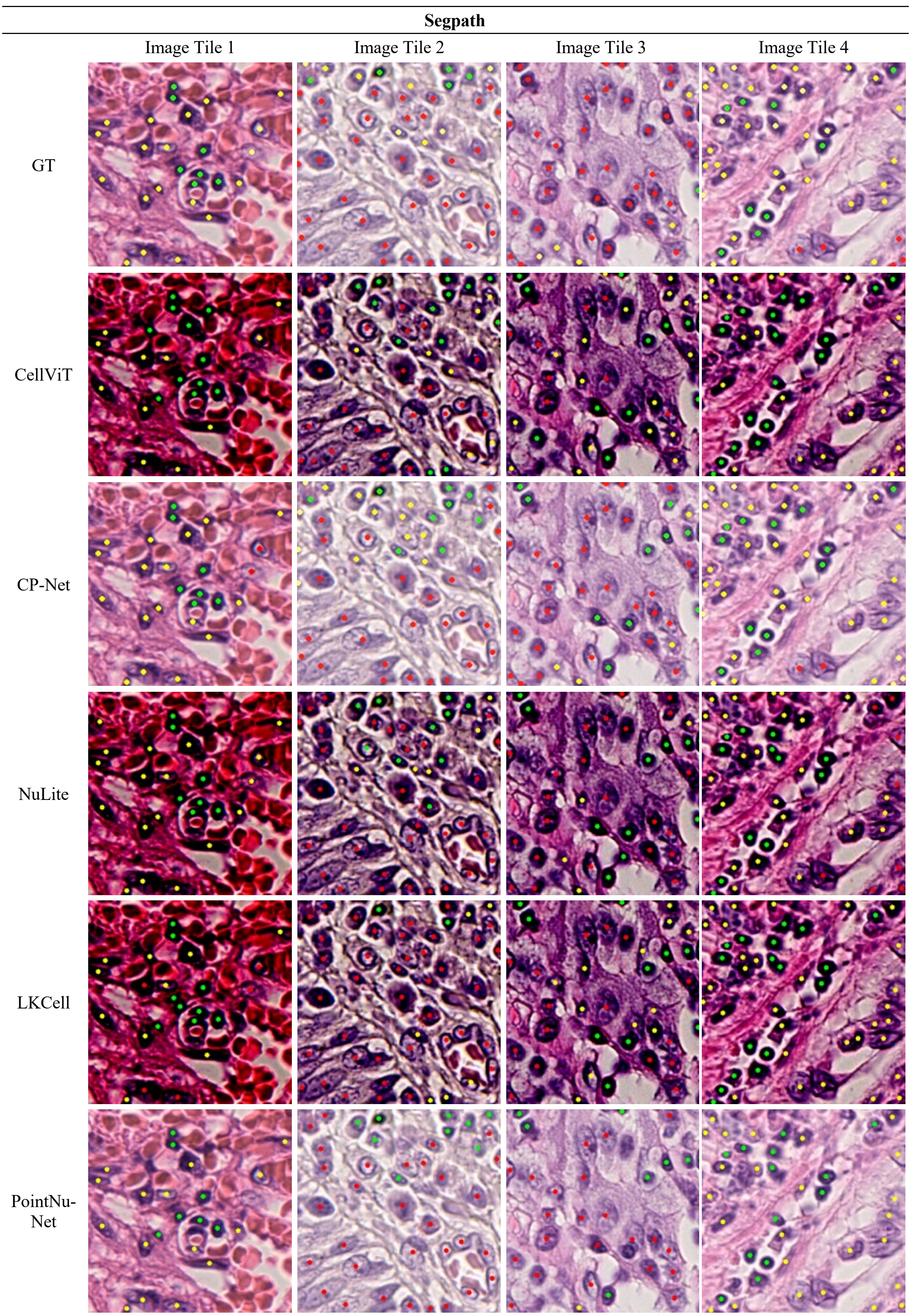

Segpath
Image Tile 1
Image Tile 2
Image Tile 3
Image Tile 4
GT
CellViT
CP-Net
NuLite
LKCell
PointNu-Net

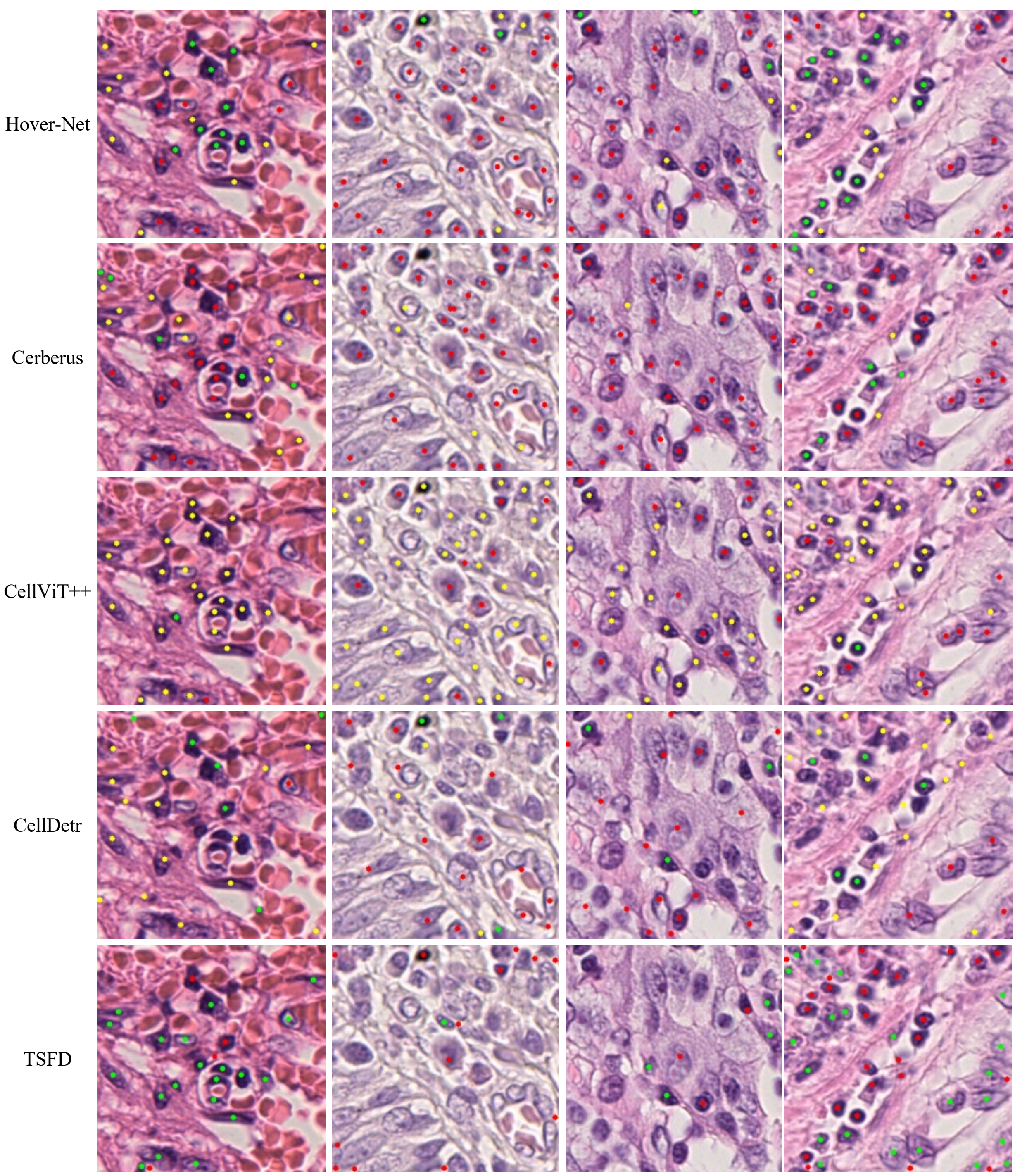
Hover-Net
Cerberus
CellViT++
CellDetr
TSFD

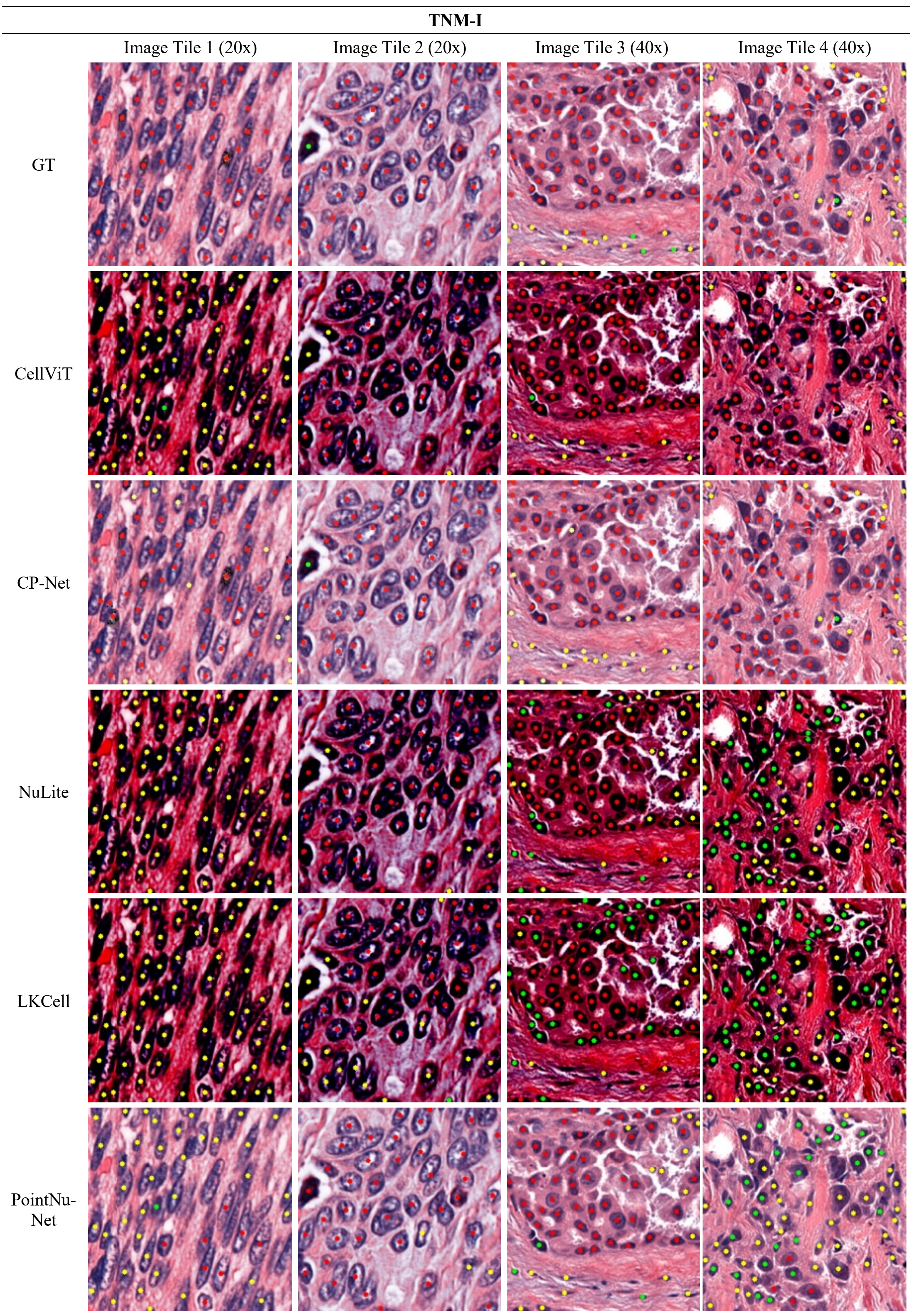
TNM-I
Image Tile 1 (20x)
Image Tile 2 (20x)
Image Tile 3 (40x)
Image Tile 4 (40x)
GT
CellViT
CP-Net
NuLite
LKCell
PointNu-Net

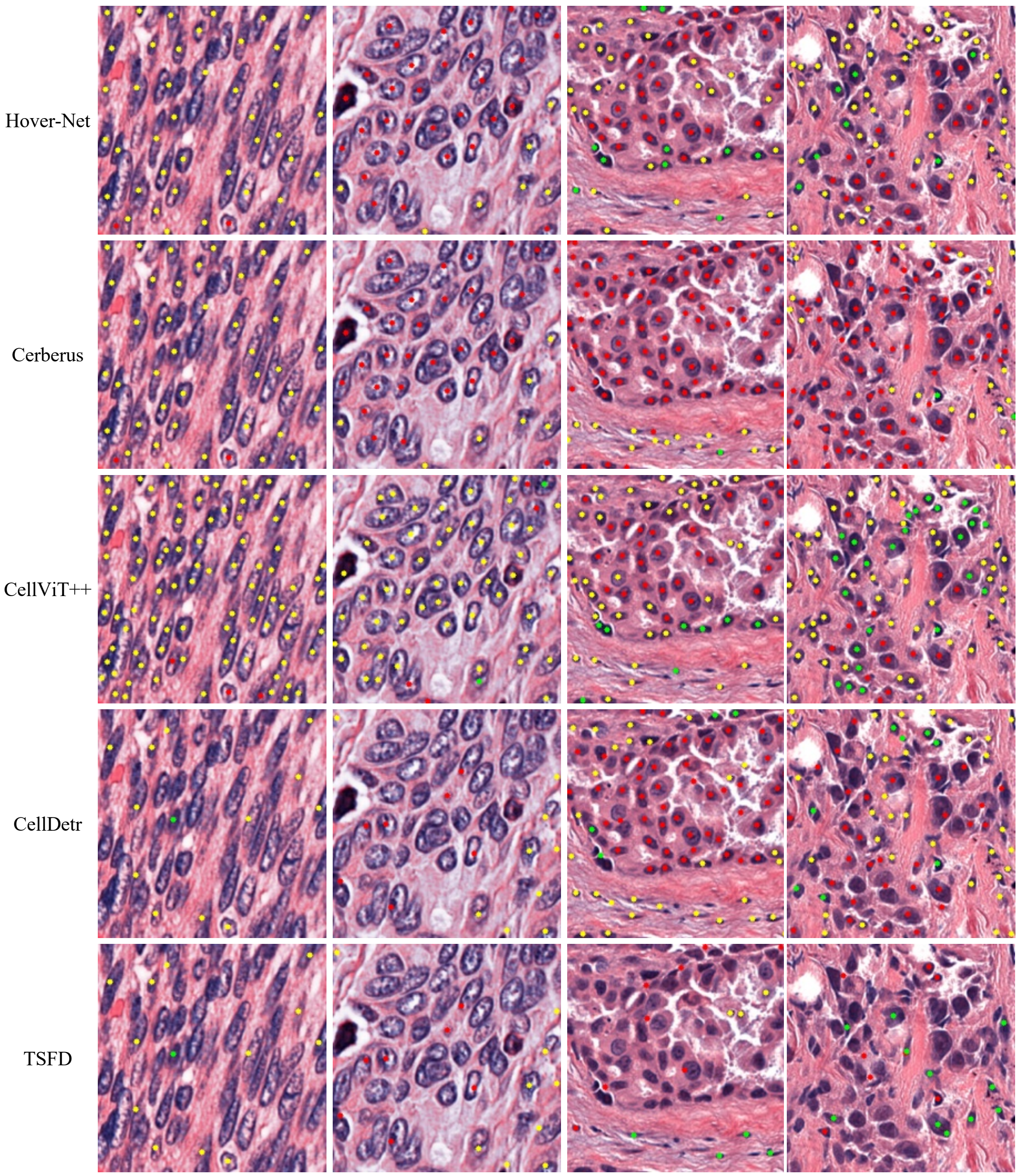

Hover-Net
Cerberus
CellViT++
CellDetr
TSFD